\documentclass[manuscript]{aastex631}
\usepackage[encapsulated]{CJK}
\usepackage{mathrsfs}
\usepackage{subfigure}
\usepackage{epstopdf}
\usepackage{amsmath}
\usepackage{longtable}
\usepackage{booktabs}
\usepackage{hyperref} \usepackage{threeparttable}
\usepackage{graphicx} 
\usepackage{overpic}
\def\etal {et al.~}

\newbox\grsign \setbox\grsign=\hbox{$>$} \newdimen\grdimen \grdimen=\ht\grsign
\newbox\laxbox \newbox\gaxbox
\setbox\gaxbox=\hbox{\raise.5ex\hbox{$>$}\llap
     {\lower.5ex\hbox{$\sim$}}}\ht1=\grdimen\dp1=0pt
\setbox\laxbox=\hbox{\raise.5ex\hbox{$<$}\llap
     {\lower.5ex\hbox{$\sim$}}}\ht2=\grdimen\dp2=0pt
\graphicspath{{figs/}}

\shortauthors{Lin \etal}

\definecolor{malachite}{rgb}{0.34, 0.7, 0.22}

\begin{document}
\begin{CJK*}{UTF8}{gbsn}

\title{Trigonometric Parallaxes of Maser Sources toward the Far Side of the Milky Way}

\correspondingauthor{Ye Xu}
\email{xuye@pmo.ac.cn}

\author{Z. H. Lin}
\affiliation{Purple Mountain Observatory, Chinese Academy of Sciences, Nanjing 210008, People's Republic of China}
\affiliation{State Key Laboratory of Radio Astronomy and Technology, Purple Mountain Observatory, Chinese Academy of Sciences, 10 Yuanhua Road, Nanjing 210023, China}

\author{Y. Xu}
\affiliation{Purple Mountain Observatory, Chinese Academy of Sciences, Nanjing 210008, People's Republic of China}
\affiliation{State Key Laboratory of Radio Astronomy and Technology, Purple Mountain Observatory, Chinese Academy of Sciences, 10 Yuanhua Road, Nanjing 210023, China}
\affiliation{University of Science and Technology of China, 96 Jinzhai Road, Hefei 230026, People's Republic of China}
\affiliation{Xinjiang Astronomical Observatory, Chinese Academy of Sciences, Urumqi, Xinjiang, 830011, People's Republic of China}

\author{M. Reid}
\affiliation{Center for Astrophysics $\vert$ Harvard \& Smithsonian, 60 Garden Street, Cambridge, MA 02138, USA}

\author{J. J. Li}
\affiliation{Purple Mountain Observatory, Chinese Academy of Sciences, Nanjing 210008, People's Republic of China}
\affiliation{State Key Laboratory of Radio Astronomy and Technology, Purple Mountain Observatory, Chinese Academy of Sciences, 10 Yuanhua Road, Nanjing 210023, China}
\affiliation{University of Science and Technology of China, 96 Jinzhai Road, Hefei 230026, People's Republic of China}

\author{S. B. Bian}
\affiliation{Purple Mountain Observatory, Chinese Academy of Sciences, Nanjing 210008, People's Republic of China}

\begin{abstract}
We present trigonometric parallaxes and proper motions for eight maser sources associated with high-mass star-forming regions. Four 22 GHz H$_2$O masers and four 6.7 GHz CH$_3$OH masers were observed over eight epochs between 2024 and 2025. Combining parallax, proper-motion, and radial-velocity information, we determine distances and spiral-arm assignments for these sources. The newly measured far-side masers extend the known distributions of the Sagittarius and Perseus arms toward Galactic longitudes as small as $\ell\sim10^\circ$. Their locations and kinematics agree well with recently updated spiral-arm models based on maser astrometry. In particular, the new measurements show that the far-side Sagittarius and Perseus arms tend to converge toward a common spiral feature in the inner Galaxy. This configuration is consistent with the bifurcation scenario proposed for the Milky Way spiral structure, although an unresolved inter-arm structure provides an alternative interpretation that cannot yet be excluded.
\end{abstract}


%
\section{Introduction}
\label{intro}
Since the first high-precision distance measurements to spiral-arm tracers were achieved through trigonometric parallaxes of masers associated with high-mass star-forming regions \citep{xu2006}, studies of the Milky Way spiral structure have entered an era of precision mapping. The Bar and Spiral Structure Legacy (BeSSeL) Survey has played a central role in this effort by systematically measuring parallaxes and proper motions of masers throughout the Galaxy \citep{brunthaler2011}. By 2019, 199 maser parallaxes had been obtained, enabling a detailed reconstruction of the large-scale spiral structure of the northern Milky Way \citep{reid2019}. In parallel, the VLBI Exploration of Radio Astrometry (VERA) project has reported parallaxes for 99 maser sources~\citep{vera2020}, of which 68 are associated with star-forming regions. A cross-match of the two samples shows that 40 of these 68 sources were already included in~\cite{reid2019}, yielding 227 distinct star-forming-region maser sources in the combined catalogs. Since then, several additional parallax measurements of masers associated with massive star-forming regions have been reported in the literature~\citep[e.g.,][]{xu2021apjs,hyland2023,sakai2023,bian2024}. To date, the total number of massive star-forming-region maser sources with accurate trigonometric parallax measurements is approaching 250; a comprehensive updated compilation will be presented by Reid et al. (in preparation). These sources are concentrated predominantly in the northern sky and close to the Galactic plane, with approximately 84\% of the sources located within $|b|<2^\circ$.

Despite this remarkable progress, our knowledge of the most distant parts of the Milky Way remains incomplete. In particular, the far side of the Galaxy, especially toward the Galactic center direction, remains sparsely sampled by trigonometric parallax measurements owing to the extreme distances and observational challenges involved. Consequently, the morphology and connectivity of spiral arms in these regions are still poorly constrained. As part of the BeSSeL Survey, we present trigonometric parallaxes and proper motions for eight maser sources associated with high-mass star-forming regions. Seven of these sources are located on the far side of the Milky Way and provide valuable new constraints on the spiral structure of the distant inner Galaxy.

\section{Observations and Data Reduction}
\label{sec2}

As part of the National Radio Astronomy Observatory's (NRAO's)\footnote{NRAO is a facility of the National Science Foundation operated under cooperative agreement by Associated Universities, Inc.} Very Long Baseline Array (VLBA) program BL312, we conducted multi-epoch astrometric observations of eight maser sources associated with high-mass star-forming regions, including four 22 GHz H$_2$O masers (G018.74$-$00.23(W), G019.27$+$00.35(W), G027.87$-$00.24(W), and G030.32$+$00.07(W)) and four 6.7 GHz CH$_3$OH masers (G011.90$-$00.10(M), G015.09$+$00.19(M), G019.61$-$00.13(M), and G025.65$+$01.05(M)). The suffixes (W) and (M) denote H$_2$O and CH$_3$OH masers, respectively. Each source was observed at eight epochs spanning approximately one year. To optimize observing efficiency, four maser sources were observed within a single 7.5 hr session, with separate sessions devoted to the H$_2$O and CH$_3$OH maser samples. The observing epochs were scheduled near the extrema of the right ascension parallax signature, where the parallax amplitude is largest, since the parallax signature in declination is generally much smaller. In total, the program required approximately 120 hr of VLBA observing time.

The correlated data were calibrated using the Astronomical Image Processing System (AIPS) together with ParselTongue scripts, following the standard procedures adopted for BeSSeL Survey astrometric observations \citep[e.g.,][]{reid2009,bian2024}. We first corrected the visibility phases for parallactic-angle variations, updated Earth-orientation parameters, and dispersive ionospheric delays estimated from global total-electron-content maps. Residual clock offsets and nondispersive tropospheric delays were determined from the geodetic-block observations and removed from the phase-referenced data. Visibility amplitudes were calibrated using the measured system temperatures and antenna gain curves, while a strong continuum calibrator was used to determine the bandpass response and remove residual delay and phase offsets among the intermediate-frequency bands. Doppler corrections were then applied to maintain a fixed LSR velocity for each maser channel over all epochs. 

Depending on the source strength, either a bright and compact maser channel or a strong and compact background quasar was adopted as the phase reference. For sources with sufficiently strong maser emission, fringe-fitting solutions derived from the reference maser channel were transferred to the remaining maser channels and the background quasars. For weaker maser sources, a suitably strong background quasar was instead used to derive the phase solutions.

For the 22 GHz H$_2$O masers, data calibration followed the standard BeSSeL Survey procedures described by \citet{reid2009}. Since ionospheric propagation delays are the dominant source of astrometric error at frequencies below $\sim10$~GHz, for the 6.7 GHz CH$_3$OH masers we used the inverse MultiView (iMV) calibration technique to acheve the highest possible astrometric precision~\citep{rioja2017,hyland2022,hyland2023}.
For G011.90$-$00.10(M), G015.09$+$00.19(M), and G019.61$-$00.13(M), nearby compact background QSOs are sparsely distributed, preventing the construction of a full iMV calibration geometry. We therefore selected two calibrator QSOs located on opposite sides of the target source, requiring the minimum distance between the target and the line connecting the two QSOs to be less than $0.5^\circ$. For G025.65$+$01.05(M), four surrounding QSOs were available and provided a complete iMV calibration geometry. The observations followed sequence (3) of \citet{hyland2022}, in which the target (T) and $N$ calibrators (C$_{1\cdots N}$) were observed in the repeating cycle T,Q$_1$,T,Q$_2$,T$\cdots$Q$_N$,T,Q$_1$,T,Q$_2\cdots$,Q$_N$,T. Since our maser targets were relatively strong, we adopted an inverse phase-referencing strategy, in which the phase solutions were derived from the maser targets and interpolated in time to the QSO scans. The observing cycle therefore intentionally begins and ends with a target scan, ensuring that each QSO scan is bracketed by target scans for temporal interpolation of the phase corrections.

For the 6.7~GHz CH$_3$OH maser observations, an additional calibration step was applied to mitigate residual direction-dependent phase errors, which are dominated by imperfectly corrected ionospheric delays at this frequency. We employed the iMV technique \citep{rioja2017,hyland2022}, using the residual phases measured toward nearby compact background QSOs to estimate the ionospheric phase gradient across the target field. The phase correction at the target position was then obtained by linearly interpolating the residual phases measured toward the two calibrators. For G025.65$+$01.05(M), four surrounding QSOs were available, allowing a complete iMV calibration geometry to be constructed. The phase corrections for this source were determined using the fitting procedure described in Section~3.2 of \citet{hyland2022}.

Images of the individual maser spectral channels were then produced with the AIPS task \texttt{IMAGR}, and the positions of the maser spots were determined by fitting two-dimensional elliptical Gaussian components with \texttt{JMFIT}. The background quasars were imaged and fitted in the same manner when their astrometric positions were required.

During the reduction of the seventh epoch of water maser, the KP antenna showed anomalous behavior that may have been caused by an instrumental or atmospheric problem, such as a clock jump or poor weather. The resulting astrometric positions showed substantially larger offsets from the expected parallax trajectory than those at the other epochs. Since including or excluding KP also produced noticeable changes in the measured positions, possibly owing to the altered $uv$ coverage, we conservatively excluded the seventh-epoch measurements from the astrometric fits.

For G011.90$-$00.10(M), the first epoch was excluded from the final fit. Owing to the absence of the NL antenna during that epoch, the synthesized beamsize was significantly larger, preventing the extended structure of the background calibrator J1819$-$2036 from being adequately resolved. This introduced a systematic offset in the measured maser position relative to the subsequent epochs.

For the astrometric analysis, we selected compact maser spots that were detected over a sufficient number of epochs and showed no evidence of strong blending or structural variability. The position of each maser spot relative to each background quasar was modeled as the sum of an annual parallax signature and a linear proper motion. Specifically, the eastward and northward position offsets were fitted simultaneously with the corresponding parallax factors calculated from the Earth's orbit, using a common parallax but independent position offsets and proper-motion components for individual maser spots. When multiple maser spots and background quasars were available, all reliable spot--quasar combinations were included in a combined solution, since the maser spots associated with a given source share the same parallax but can have different internal motions \citep{reid2009,bian2024}. Separate error floors were added in quadrature to the formal positional uncertainties in the eastward and northward directions and adjusted until the reduced $\chi^{2}$ values were close to unity, thereby accounting for residual atmospheric delays, unresolved quasar structure, and other systematic astrometric errors. Because such systematic errors can be correlated among maser spots observed at the same epoch, the formal uncertainty of a combined parallax solution was conservatively increased by the square root of the number of maser spots used in the fit. 
The resulting parallaxes and proper motions are summarized in Table~\ref{tabs1}. Additional details of observing epochs, source/calibrator information, and individual astrometric fits are provided in Appendix~\ref{SecA}.

\vspace{-0.5cm}
\begin{deluxetable}{ccccccc}[!ht]
\tablecolumns{6}
\tablecaption{Parallaxes and Proper Motions\label{tabs1}}
\tablehead{
\colhead{Source} & \colhead{$V^{*}_{\rm LSR}$} & \colhead{$\mu_x$} & \colhead{$\mu_y$} & \colhead{$\pi$} & \colhead{$D_{\rm 3DKD}$} & \colhead{$D$} \\ 
\colhead{} & \colhead{km s$^{-1}$} & \colhead{(mas yr$^{-1}$)} & \colhead{(mas yr$^{-1}$)} & \colhead{(mas)} & \colhead{(kpc)}  & \colhead{(kpc)} 
}
\startdata
G018.74$-$00.23(W) & $38\pm5$\tablenotemark{a} & $-3.38\pm0.03$ & $-6.46\pm0.15$ & $0.081\pm0.010$ & $12.80\pm0.76$ & $12.74\pm0.68$ \\
G019.27$+$00.35(W) & $16\pm5$\tablenotemark{a} & $-3.05\pm0.04$ & $-6.59\pm0.20$ & $0.072\pm0.013$ & $13.06\pm0.90$ & $13.18\pm0.82$ \\
G027.87$-$00.24(W) & $20\pm5$\tablenotemark{b} & $-2.97\pm0.04$ & $-5.84\pm0.16$ & $0.096\pm0.012$ & $13.34\pm0.86$ & $12.66\pm0.99$ \\
G030.32$+$00.07(W) & $45\pm5$\tablenotemark{c} & $-3.10\pm0.03$ & $-6.19\pm0.23$ & $0.086\pm0.011$ & $11.64\pm1.09$ & $11.69\pm0.84$\\
\hline
G011.90$-$00.10(M) & $34\pm5$ & $-3.75\pm0.09$ & $-6.15\pm0.47$ & $0.133\pm0.030$ & $13.44\pm1.13$ & $12.88\pm1.10$\\
G015.09$+$00.19(M) & $26\pm5$ & $-3.05\pm0.16$ & $-5.46\pm0.34$ & $0.034\pm0.055$ & $14.97\pm1.43$ & $15.09\pm1.43$\\
G019.61$-$00.13(M) & $55\pm5$ & $-3.72\pm0.19$ & $-6.77\pm0.53$ & $0.038\pm0.066$ & $11.84\pm0.99$ & $11.97\pm0.99$\\
G025.65$+$01.05(M) & $42\pm5$ & $+0.55\pm0.04$ & $-2.46\pm0.12$ & $0.359\pm0.016$ & $2.64\pm0.75$ & $2.79\pm0.12$\\
\enddata
\tablecomments{(W) and (M) denote 22 GHz H$_2$O and 6.7 GHz CH$_3$OH masers, respectively. a: \cite{titmarsh2014}; b: \cite{breen2015}; c: \cite{bartkiewicz2011}.}
\end{deluxetable}

For the H$_2$O maser sources, the systemic velocity, $V^{*}_{\rm LSR}$, was adopted from the associated 6.7 GHz CH$_3$OH masers, whose velocities are generally considered to provide a more reliable tracer of the systemic motion of the central exciting source \citep{bartkiewicz2011,breen2015,titmarsh2014}. Since 6.7 GHz CH$3$OH masers typically differ from the systemic velocity of the exciting source by about 5 km s$^{-1}$ \citep{moscadelli2002}, we adopted 5 km s$^{-1}$ as the uncertainty floor for $V^{*}_{\rm LSR}$.


\section{Results}
\label{sec3}
\subsection{Distance}

We estimated the source distances by combining the probability density functions derived from the trigonometric parallax and the three-dimensional kinematic distance ($D_{\rm 3DKD}$) method of \citet{reid2022}. Use notation similar to that in~\citep{reid2022}, we calculated the likelihood function for
the parallax component ${\rm Prob}(d\mid \varpi, \sigma_\varpi)$, LSR velocity component ${\rm Prob}(d\mid V_{\rm LSR}, \sigma_{\rm LSR}, RC)$, Galactic longitude proper-motions ${\rm Prob}(d\mid \mu_l^{*}, \sigma_{\mu_l^{*}}, RC)$ and latitude proper-motions ${\rm Prob}(d\mid \mu_b, \sigma_{\mu_b}, RC)$ as follow,
\begin{align}
{\rm Prob}(d\mid \varpi, \sigma_\varpi)  = 
\frac{1}{\sqrt{2\pi}\,\sigma_\varpi}
\exp\left[
-\frac{\left(\varpi-1/d\right)^2}
{2\sigma_\varpi^2}
\right], & \\
{\rm Prob}(d\mid i,\sigma_i,RC)  = 
\frac{1}{\sqrt{2\pi}\,\sigma_i}
\exp\left[
-\frac{\left(x_{i}-x_{i,{\rm RC}}(d)\right)^2}
{2\sigma_i^2}
\right], &
\qquad
i=V_{\rm LSR},\mu_l^{*},\mu_b,
\end{align}
where $d$ is in kpc, and $x_{i,{\rm RC}}$ is the value predicted by the adopted Galactic rotation curve (RC) model, calculated following the procedure described in Appendix~A of \citet{hyland2026}.
For the parallax likelihood, $\sigma_\varpi$ was taken to be the measured parallax uncertainty. For each of the three kinematic constraints, we included an additional velocity dispersion of $\sigma_{\rm disp}=10$~km~s$^{-1}$ in quadrature with the observational uncertainty to account for non-circular source motions and imperfections in the Galactic rotation model. Thus, for $V_{\rm LSR}$,
\begin{equation}
\sigma_{V}^{2}
=
\sigma_{V,{\rm obs}}^{2}
+
\sigma_{\rm disp}^{2},
\end{equation}
whereas, for the proper-motion components, the additional velocity dispersion was converted to an angular-motion uncertainty at each trial distance:
\begin{equation}
\sigma_{\mu_j}^{2}(d)
=
\sigma_{\mu_j,{\rm obs}}^{2}
+
\left(\frac{\sigma_{\rm disp}}{4.74\,d}\right)^2,
\qquad j=l^{*},b,
\end{equation}
where factor 4.74 converts proper motion in mas yr$^{-1}$ at a distance in kpc into transverse velocity in km s$^{-1}$.

After normalization over the adopted distance interval, each kinematic likelihood was assigned a reliability weight of $w_{\rm kin}=0.85$. Operationally, this was implemented as a mixture of the calculated kinematic probability density and a uniform probability density containing the remaining probability of 0.15:
\begin{equation}
\widetilde{\rm P}(d\mid i)
=
w_{\rm kin}{\rm P}(d\mid i, \sigma_i, RC)
+
\left(1-w_{\rm kin}\right)\mathcal{U}(d),
\qquad
i=V_{\rm LSR},\mu_l^{*},\mu_b,
\end{equation}
where $\mathcal{U}(d)=1/(d_{\max}-d_{\min})$ within the sampled distance range. This mixture reduces the likelihood that a single discrepant kinematic observable, caused for example by a peculiar source motion, completely suppresses an otherwise plausible distance solution. 

The parallax likelihood was assigned a weight of unity and therefore contained no uniform component. The final distance probability density was obtained by multiplying the parallax and three weighted kinematic likelihoods and subsequently normalizing the product:
\begin{equation}
{\rm P}(d)
=
\frac{
{\rm P}(d\mid\varpi)
\widetilde{{\rm P}}(d\mid V_{\rm LSR})
\widetilde{{\rm P}}(d\mid{\mu_l^{*}})
\widetilde{{\rm P}}(d\mid{\mu_b})}
{\displaystyle
\int_{d_{\min}}^{d_{\max}}
{\rm P}(d\mid\varpi)
\widetilde{\rm P}(d\mid V_{\rm LSR})
\widetilde{\rm P}(d\mid\mu_l^{*})
\widetilde{\rm P}(d\mid\mu_b){\rm d}d}.
\end{equation}

After multiplication and normalization of the weighted probability distributions, we obtained the final likelihood probability density as a function of distance for each maser source. A Gaussian function was fitted to the dominant peak of the likelihood distribution, and the fitted peak position (most probable distance, $D$) and standard deviation were adopted as the source distance and its $1\sigma$ uncertainty, respectively.

\begin{figure}[!ht]
    \centering
    \includegraphics[width=0.38\linewidth]{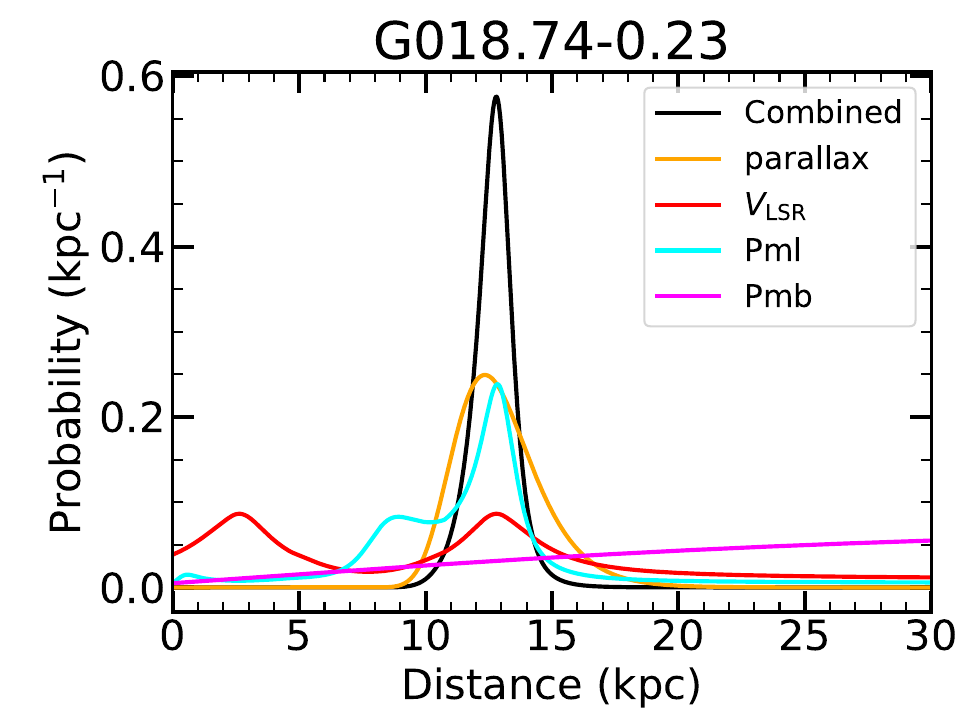}
    \includegraphics[width=0.38\linewidth]{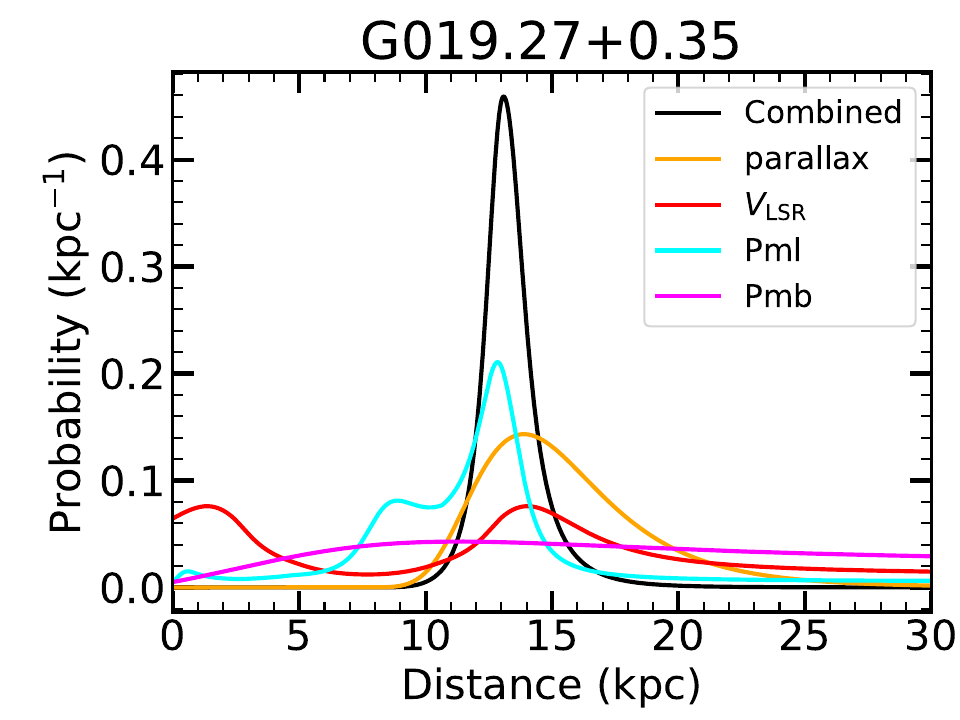}
    \includegraphics[width=0.38\linewidth]{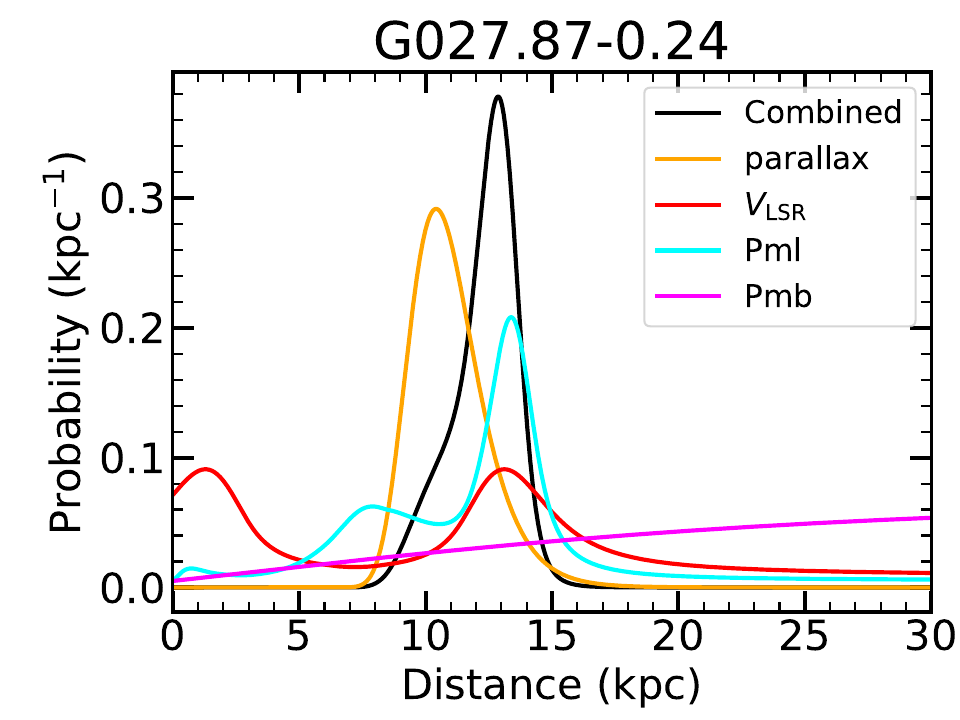}
    \includegraphics[width=0.38\linewidth]{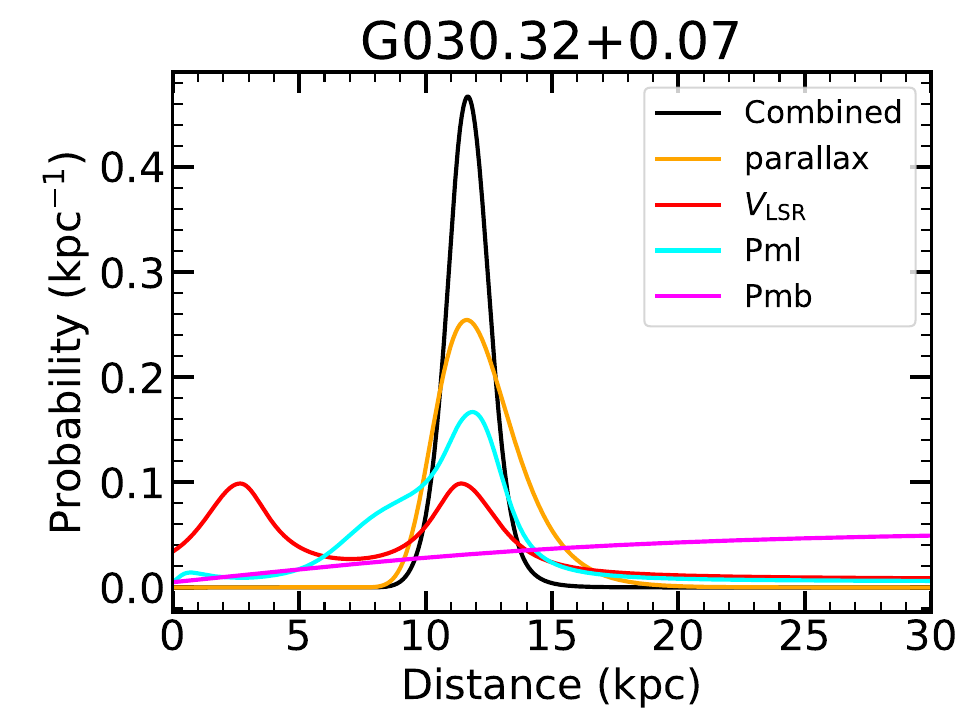}
    \includegraphics[width=0.38\linewidth]{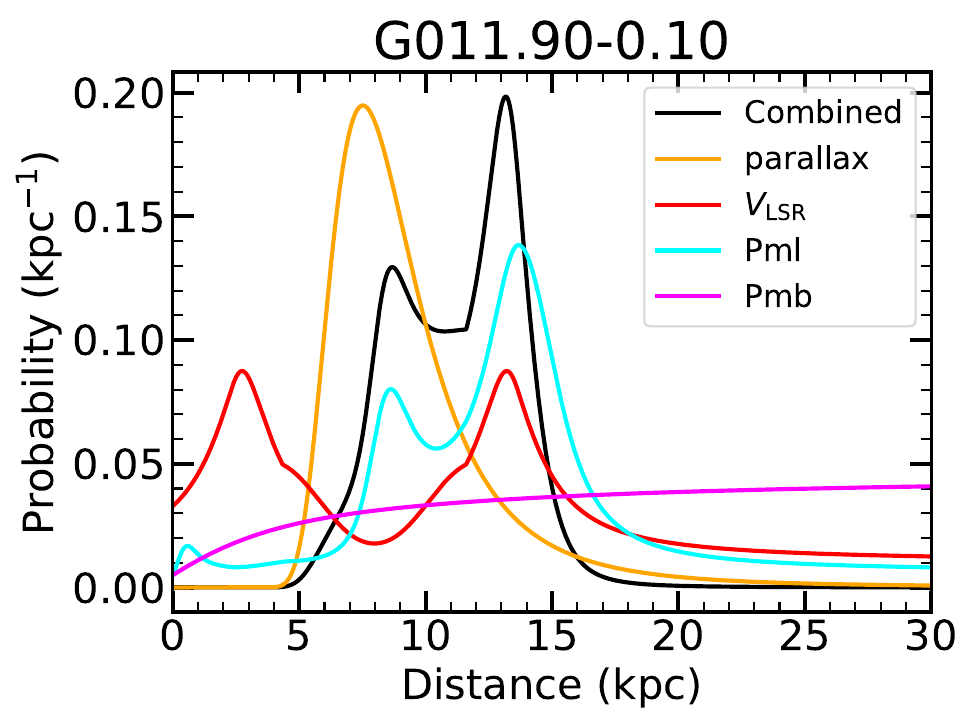}
    \includegraphics[width=0.38\linewidth]{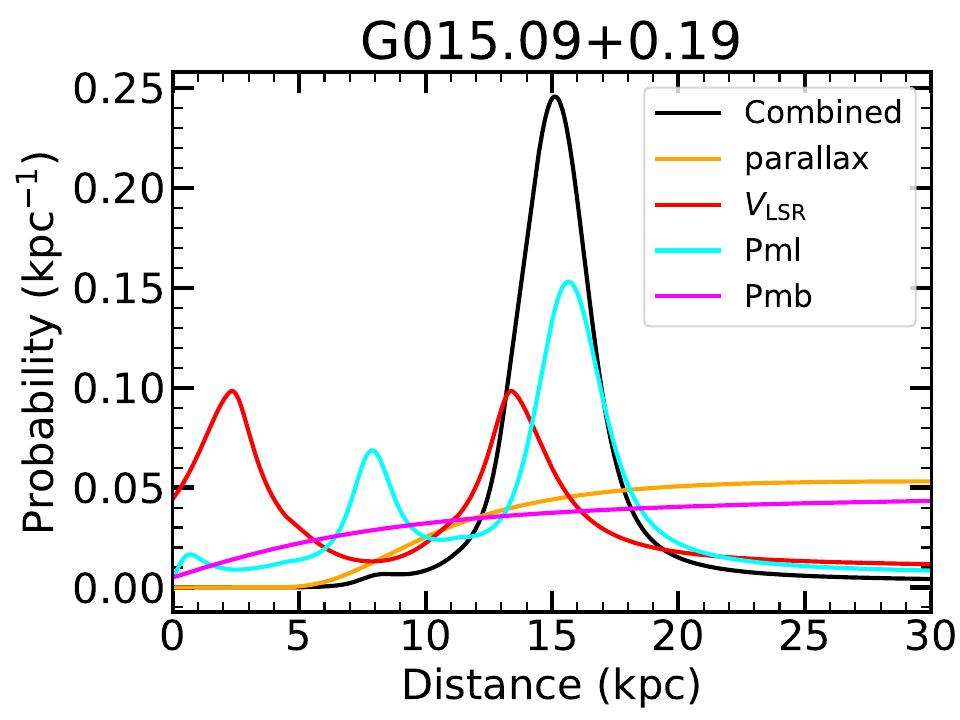}
    \includegraphics[width=0.38\linewidth]{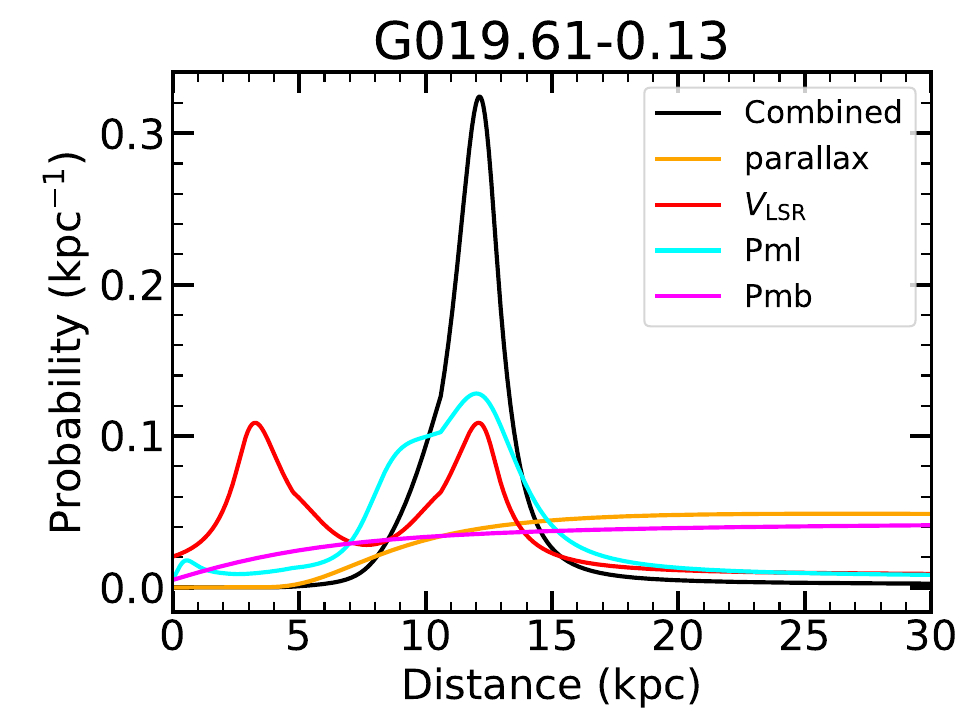}
    \includegraphics[width=0.38\linewidth]{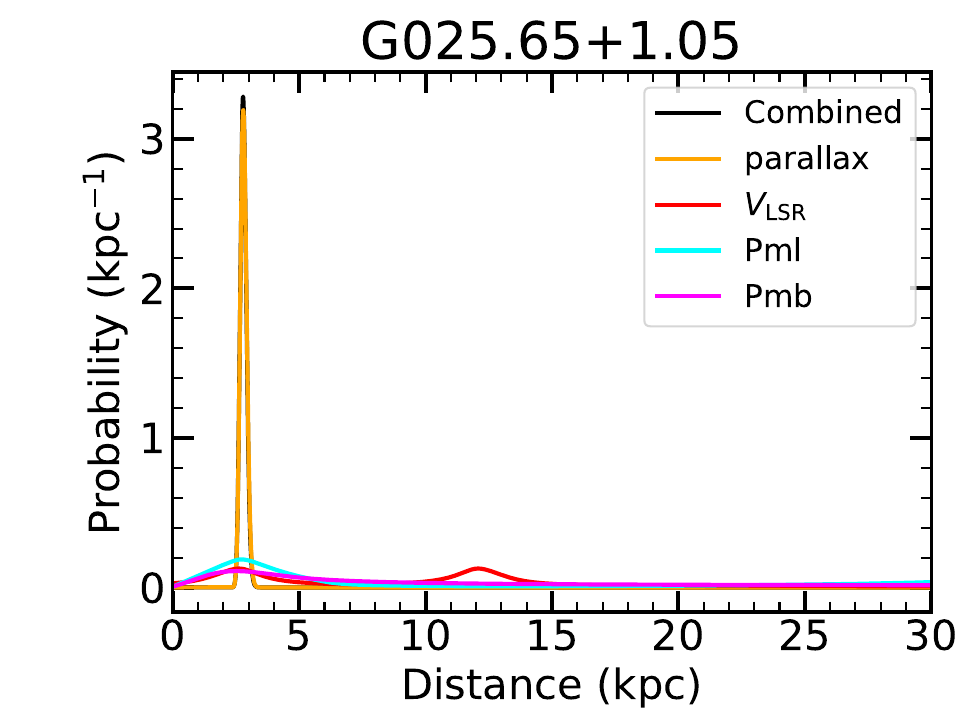}
    \caption{Distance probability distributions for the eight maser sources. The red, cyan, magenta, and orange curves represent the probability density functions derived from the radial velocity, proper motion in Galactic longitude, proper motion in Galactic latitude, and trigonometric parallax, respectively. The black curve shows the combined likelihood probability distribution obtained by multiplying all individual probability density functions.}
    \label{figs:distance}
\end{figure}

Figure~\ref{figs:distance} presents the resulting distance probability distributions. All sources, except G025.65$+$01.05(M), are found to lie on the far side of the Milky Way.

The parallax, proper-motion, and radial-velocity measurements consistently place G018.74$-$00.23(W), G019.27$+$00.35(W), and G030.32$+$00.07(W) on the far side of the Milky Way, with adopted distances of $12.74\pm0.68$, $13.18\pm0.82$, and $11.69\pm0.84$ kpc, respectively. For G027.87$-$00.24(W), the trigonometric parallax alone implies a distance of $10.4\pm1.3$~kpc, whereas the kinematic constraints from the radial velocity and proper motions favor a far-side distance of $13.34\pm0.86$~kpc. The two estimates differ by about $1.9\sigma$ when the uncertainties of both measurements are taken into account. Combining the constraints from the parallax, radial velocity, and proper motions, we adopt a final distance of $12.66\pm0.99$ kpc for this source.

For G011.90$-$00.10(M), G015.09$+$00.19(M), and G019.61$-$00.13(M), the relatively large parallax uncertainties are most likely dominated by residual ionospheric delays at 6.7~GHz. 
Although the two-calibrator iMV correction removes a substantial fraction of the direction-dependent phase errors, the maser positions do not lie exactly on the lines connecting the two background QSOs. Therefore, residual ionospheric phase gradients, particularly those with components perpendicular to the calibrator lines, can remain uncorrected. \citet{reid2017} showed that epoch-dependent position gradients of order $0.1$~mas~deg$^{-1}$ with varying orientations can introduce significant systematic errors in 6.7~GHz maser astrometry. Such residual gradients may therefore account for the relatively large astrometric uncertainties of these three sources. In contrast, the four QSOs surrounding G025.65$+$01.05(M) provide two-dimensional constraints on the phase gradient and yield a substantially more precise parallax measurement.

Although the parallaxes of G011.90$-$00.10(M), G015.09$+$00.19(M), and G019.61$-$00.13(M) are less precise than that of G025.65$+$01.05(M), their conservative $3\sigma$ distance lower limits still exceed 4~kpc. These limits are beyond the corresponding near kinematic distances of $\sim$3~kpc, thereby excluding the near-distance solutions. In addition, while the distance likelihoods derived separately from $V_{\rm LSR}$ and the proper motions can admit multiple solutions, their dominant probability peaks are consistent with the distant solutions indicated by the parallaxes. By combining the parallax and three-dimensional kinematic likelihoods, we derive distances of $12.88\pm1.10$, $15.09\pm1.43$, and $11.97\pm0.99$~kpc for G011.90$-$00.10(M), G015.09$+$00.19(M), and G019.61$-$00.13(M), respectively.

G025.65$+$01.05(M), also known as IRAS 18316$-$0602 or RAFGL 7009S, is an active high-mass star-forming region associated with strong H$_2$O, CH$_3$OH, and OH maser emission \citep{bayandina2019}. Owing to its recurrent H$_2$O maser superbursts, during which the flux density can reach tens of thousands of Jy, this source has become an important target for studies of maser burst mechanisms \citep[e.g.,][]{volvach2019,burns2020}. 
Previous distance estimates for G025.65$+$01.05 were based primarily on its line-of-sight velocity, $V_{\rm LSR}$, together with an assumed Galactic rotation model, and were therefore subject to the classical near--far kinematic distance ambiguity in the first Galactic quadrant. Molecular-line studies reported near kinematic distances of 3.17 and 2.7~kpc \citep{molinari1996,sunada2007}, whereas the H{\sc i} self-absorption analysis of \citet{green2011} assigned the source to the far kinematic distance of 12.5~kpc. In contrast, our analysis incorporates not only $V_{\rm LSR}$ but also the two transverse components of motion. In particular, the measured Galactic-longitude proper motion, $\mu_l^{*}$, strongly favors the near-distance solution, while the trigonometric parallax provides a direct geometric constraint. The measured parallax corresponds to a distance of $2.76\pm0.12$~kpc, confirming that G025.65$+$01.05 lies at the near distance and ruling out the previously proposed 12.5~kpc far-distance solution.

The luminosity inferred for the central exciting source of G025.65$+$01.05 depends strongly on the adopted distance. At a near distance of $\sim3$~kpc, the source was suggested to be excited by a B1-type star, whereas the previously proposed far distance of 12.5~kpc would imply a bolometric luminosity comparable to that of an O4 star \citep{volvach2019}. Using our accurately determined distance of $2.76\pm0.12$~kpc, the bolometric luminosity derived from the SED modeling of \citet{mookerjea1999} can be rescaled from $2.54\times10^{4}~L_{\odot}$ at 3.3~kpc to approximately $1.8\times10^{4}~L_{\odot}$. Under a single-star interpretation, this luminosity is consistent with an early-B-type massive star, removing the need for an extremely luminous O-type exciting source.

\subsection{Spiral Structure}

Since this study focuses on maser sources located on the far side of the Milky Way, G025.65$+$01.05(M) is not discussed further. Galactic spiral arms appear as coherent large-scale structures in longitude, latitude, and line-of-sight velocity space. In particular, H{\sc i} and CO observations delineate approximately continuous spiral-arm features in longitude--velocity diagrams \citep{weaver1970,cohen1980,reid2016}. We therefore determined the spiral-arm association of each maser source by quantitatively comparing its observed $(l,b,V_{\rm LSR})$ coordinates with the corresponding $(l,b,V_{\rm LSR})$ traces of the candidate arms.

For each candidate arm, we adopted the $\ell$--$b$--$V_{\rm LSR}$ parameters from \citet{reid2016,reid2019} and calculated the latitude and velocity residuals
\begin{equation}
\Delta b_i=b_{\rm s}-b_i(l_{\rm s})~{\rm and}~\Delta V_i=V_{{\rm LSR},{\rm s}}-V_i(l_{\rm s}),
\end{equation}
Here, $_i$ represents the $i$th spiral arm and $_s$ represents the maser source.

The physical arm width was converted to an angular width at the model distance according to
\begin{equation}
\theta_{W,i}
=
\tan^{-1}\left[\frac{W_i}{d_i(l_{\rm s})}\right],
\end{equation}
and assume that $W_i$ is all 0.3 kpc for each arm.

The latitude likelihood was evaluated using a Gaussian probability density,
\begin{equation}
\mathcal{L}_{b,i}
=
\frac{1}{\sqrt{2\pi}\theta_{W,i}}
\exp\left[
-\frac{\Delta b_i^{2}}{2\theta_{W,i}^{2}}
\right].
\end{equation}

Similarly, the effective velocity uncertainty was defined as
\begin{equation}
\sigma_{V,i}^{2}
=
\sigma_{V,{\rm s}}^{2}
+
\sigma_{\rm pec}^{2},
\end{equation}
where $\sigma_{V,{\rm s}}$ is the uncertainty in the measured systemic velocity and $\sigma_{\rm pec}=10$~km~s$^{-1}$ accounts for typical non-circular and internal motions of high-mass star-forming regions relative to an idealized spiral-arm velocity trace. The corresponding velocity likelihood was
\begin{equation}
\mathcal{L}_{V,i}
=
\frac{1}{\sqrt{2\pi}\sigma_{V,i}}
\exp\left[
-\frac{\Delta V_i^{2}}{2\sigma_{V,i}^{2}}
\right].
\end{equation}
Assuming that the latitude and velocity residuals provide independent constraints at a fixed longitude, the joint likelihood for association with arm $i$ was calculated as
\begin{equation}
\mathcal{L}_i
=
\mathcal{L}_{b,i}\mathcal{L}_{V,i}.
\end{equation}
We adopted equal prior probabilities for all candidate arms so that the assignments were not influenced by previously published source labels. The normalized arm-association probability was therefore
\begin{equation}
P({\rm arm}\mid l_{\rm s},b_{\rm s},V_{{\rm LSR},{\rm s}})
=
\frac{\mathcal{L}_i}
{\sum_{j}\mathcal{L}_j}.
\end{equation}

The final spiral-arm assignments for the seven maser sources are summarized in Table~\ref{tab:spiral_assign}. Among them, G018.74$-$00.23(W), G030.32$+$00.07(W), G011.90$-$00.10(M), and G019.61$-$00.13(M) have the highest probabilities of being associated with the Sagittarius Arm, whereas G019.27$+$00.35(W), G027.87$-$00.24(W), and G015.09$+$00.19(M) are most likely associated with the Perseus Arm.

\begin{deluxetable}{ccccc}[!ht]
\tablecolumns{5}
\tablecaption{Spiral-arm assignments\label{tab:spiral_assign}}
\tablehead{
\colhead{Source} & \colhead{Arm}& \colhead{Prob}  & \colhead{Arm}& \colhead{Prob}
}
\startdata
        G018.74$-$00.23(W) &  Sagittarius & 0.59 & Perseus & 0.39\\
        G019.27$+$00.35(W) &  Perseus & 0.54 & Outer & 0.28\\
        G027.87$-$00.24(W) &  Perseus & 0.80 & Sagittarius & 0.12\\
        G030.32$+$00.07(W) &  Sagittarius & 0.73 & Perseus & 0.27 \\
        G011.90$-$00.10(M) &  Sagittarius & 0.58 & Perseus & 0.41\\
        G015.09$+$00.19(M) &  Perseus & 0.56 & Sagittarius & 0.41\\
        G019.61$-$00.13(M) &  Sagittarius & 0.84 & 
        Perseus &  0.12\\
\enddata
\tablecomments{Spiral-arm association probabilities were calculated for the relevant \citet{reid2019} spiral-arm models on the far side of the first Galactic quadrant for each of the seven maser sources, based on their $\ell$--$b$--$V_{\rm LSR}$ distributions. Only the most probable and second-most probable arm associations, together with their corresponding probabilities, are listed here.}
\end{deluxetable}

\begin{figure}[!ht]
    \centering
    \includegraphics[width=0.90\linewidth]{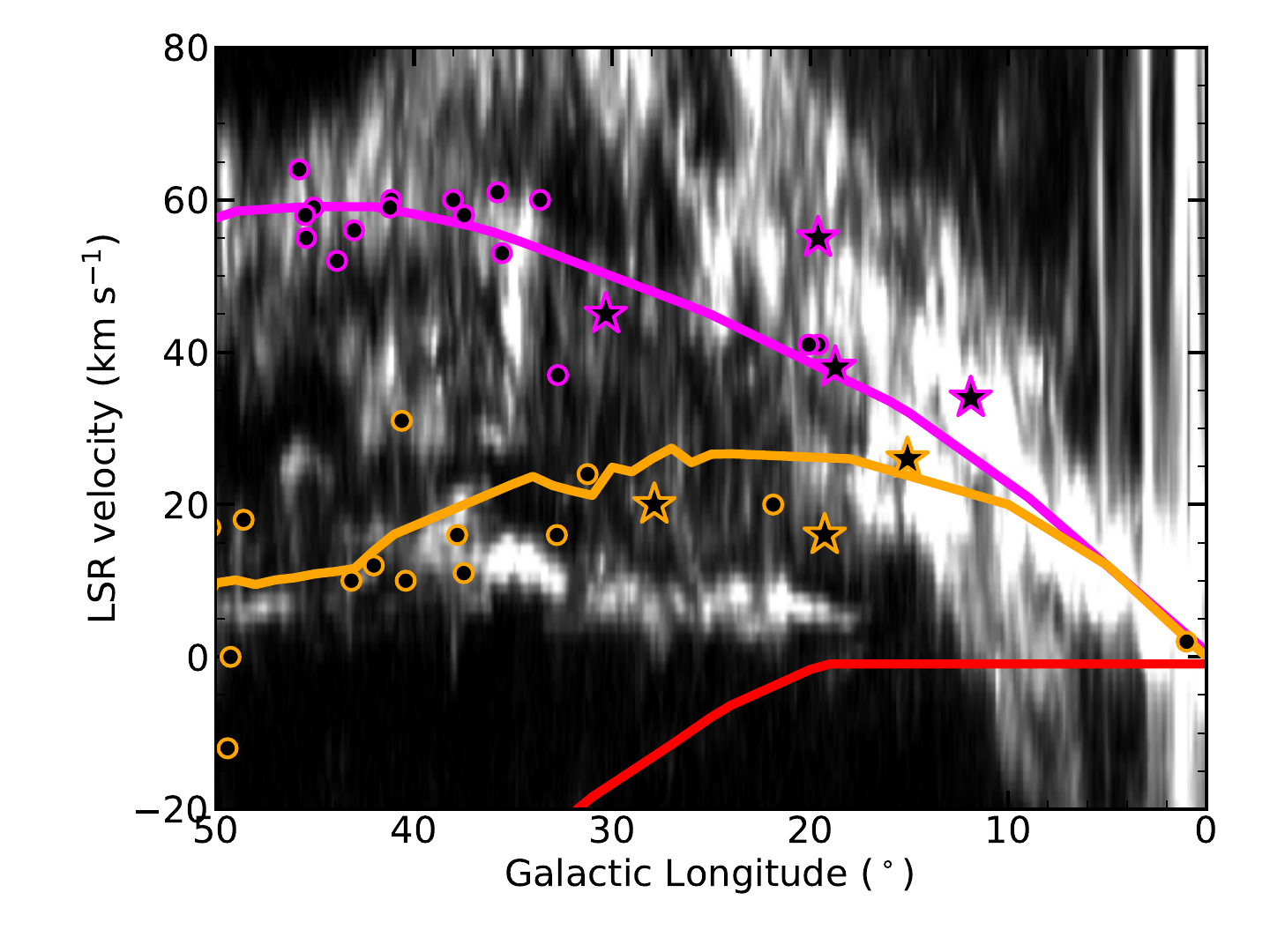}
    \caption{Location of masers superposed on a CO (l, v) diagram from~\cite{dame2001}; the far side Sagittarius (magenta), Perseus (orange), and Outer (red) traces are from~\cite{reid2016,reid2019}. Black dots with magenta/orange edges indicate previously published sources assigned to the far portion of the Sagittarius/Perseus arm~\citep{reid2019,bian2024,hyland2026}. Seven black pentagrams with magenta/orange edges indicate sources assigned to the far portion of the Sagittarius/Perseus arm in this work.}
    \label{figs:assign}
\end{figure}

Figure~\ref{figs:assign} shows the distribution of these sources in $\ell$--$b$--$V_{\rm LSR}$ space. In addition, combining our measurements with the previously published maser astrometry and compilations for these two arms \citep{bian2024,hyland2026}, we also plot in Figure~\ref{figs:assign} the far-side maser sources in the Sagittarius and Perseus arms for which accurate astrometric measurements are currently available (as shown in Table~\ref{tab:masers_all}).

\begin{figure}[!ht]
    \centering
    \includegraphics[width=0.70\linewidth,trim=3.5cm 3cm 1.5cm 1.5cm,clip]{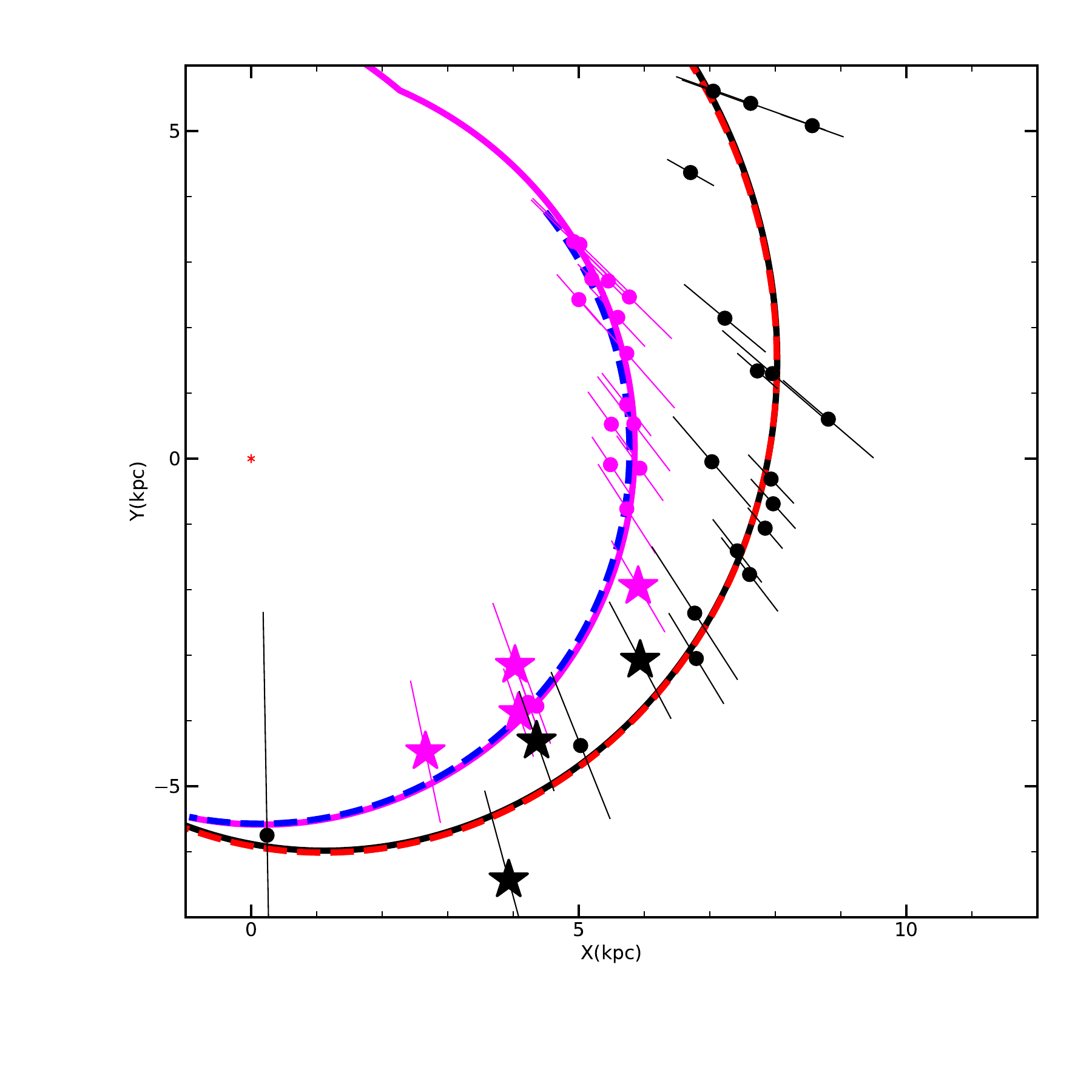}
    \caption{Plan view of the Milky Way as seen from the north Galactic pole. The Galactic center is at (0, 0) kpc and the Sun is at (0, 8.15) kpc. Solid dots indicate locations of the previously published maser sources. Pentagrams mark the locations of the seven maser sources measured in this work. Distance uncertainties are indicated by the inverse size of the symbols as given in the legend at the lower left. The solid lines show the two-segment Sagittarius Arm model \citep[magenta;][]{bian2024} and the Perseus Arm model \citep[black;][]{hyland2026}, while the blue and red dash lines represent the newly fitted Sagittarius and Perseus Arm models, respectively.}
    \label{figs:plan_view}
\end{figure}

The distribution of these sources in the Galactic $X$--$Y$ plane is shown in Figure~\ref{figs:plan_view}. For comparison, we also plot the updated Sagittarius Arm model refitted by \citet{bian2024} and the revised Perseus Arm model presented by \citet{hyland2026}. The positions of all seven sources are consistent with these updated spiral-arm models. In particular, the newly measured masers associated with the Sagittarius Arm extend its continuous tracing inward to $\ell\sim10^\circ$.

With the addition of recent far-side maser parallax measurements, the Sagittarius and Perseus arms show a tendency to converge on the far side of the Milky Way. This geometry is consistent with a scenario in which the two arms connect or merge toward a common spiral feature, and is also compatible with the inner two-arm and outer multi-arm bifurcation model proposed by \citet{xu2023}. However, the present data do not uniquely distinguish this interpretation from other possible configurations. In particular, an as-yet unresolved inter-arm structure between the Sagittarius and Perseus arms could also account for some of the measured positions, especially those of G011.90$-$00.10(M) and G015.09$+$00.19(M), whose arm-association probabilities do not strongly exclude the alternative arm. We therefore regard the bifurcation interpretation as one possible explanation rather than a unique conclusion. Additional parallax measurements sampling the region at smaller Galactic longitudes and between the currently traced arms will be required to distinguish these scenarios.

All available maser data points associated with the far-side Sagittarius Arm and the first-quadrant Perseus Arm were fitted with a log-periodic spiral model~\citep{reid2019,xu2023,bian2024,hyland2026}, and the resulting spiral-arm parameters are listed in Table~\ref{tab:arm_fit}. Since the ``kink'' locations identified in previous studies \citep{reid2019,bian2024,hyland2026} lie outside the arm segments considered here, a single pitch angle was adopted for each fit. The resulting spiral-arm models are also shown in Figure~\ref{figs:plan_view} and are in good agreement with the previous fits of the Sagittarius and Perseus Arms presented by \citet{bian2024} and \citet{hyland2026}, respectively.

\begin{deluxetable*}{lcccc}[!h]
\tablecaption{Arm Fitting Results\label{tab:arm_fit}}
\tablewidth{0pt}
\tablehead{
\colhead{Reference} &
\colhead{$\beta_{\rm kink}$} &
\colhead{$R_{\rm kink}$} &
\colhead{$\psi_{<}$} &
\colhead{$\psi_{>}$}\\
\colhead{} &
\colhead{(deg)} &
\colhead{(kpc)} &
\colhead{(deg)} &
\colhead{(deg)} 
}
\startdata
\multicolumn{5}{c}{Sagittarius arm}\\
\hline
\cite{bian2024} & 22 & $6.06\pm0.06$ & $18.4\pm1.4$ & $1.7\pm1.0$\\
This work & 22 & $5.93\pm0.10$ & ... & $1.3\pm2.0$\\
\hline
\multicolumn{5}{c}{Perseus arm}
\\
\hline
\cite{hyland2026} & 40 & $9.29\pm0.10$ & $5.9\pm1.2$ & $10.6\pm1.0$\\
This work & 40 & $9.26\pm0.12$ & ... & $10.3\pm1.7$
\enddata
\end{deluxetable*}

According to the current spiral-arm model, G001.00$-$00.23 is assigned to the Perseus arm \citep{hyland2026}. However, in this region the separation between the Sagittarius and Perseus arms is already smaller than their characteristic arm widths, such that the spatial loci of the two arms become effectively degenerate. If the current spiral-arm model is correct, the two arms may already have converged or begun to merge in this region, although an unresolved intermediate structure cannot be excluded. It is therefore difficult to uniquely assign G001.00$-$00.23 to either arm based on the currently available observations. Additional trigonometric parallax measurements of maser sources at Galactic longitudes $\ell<10^\circ$ will be required to determine whether the observed distribution is best described by arm convergence, or an additional intermediate structure.

\section{summary}
\label{sec4}

In this paper, we present VLBA trigonometric parallax and proper-motion measurements for eight maser sources associated with high-mass star-forming regions. The main results are summarized as follows.

1. We measured parallaxes and proper motions for four 22 GHz H$_2$O masers and four 6.7 GHz CH$_3$OH masers. Combining the parallax, proper-motion, and radial-velocity information, we determined reliable distances for all eight sources. Seven sources are located on the far side of the Milky Way at distances of $\sim11$--15 kpc.

2. Based on the measured distances, $\ell-b-v$ distributions, and the latest maser parallax measurements, we assign the seven far-side maser sources to spiral arms. Four sources are associated with the far-side Sagittarius arm, while the remaining three belong to the far-side Perseus arm.

3. Together with recent VLBI parallax measurements, our results show that the far-side Sagittarius and Perseus arms tend to converge toward a common spiral feature in the inner Galaxy. This configuration is consistent with the bifurcation model proposed by \citet{xu2023}, but alternative explanations, including an unresolved inter-arm structure, cannot yet be excluded. Future parallax measurements at $\ell<10^\circ$ and in the region between the two arms will be important for distinguishing these possibilities.

\begin{acknowledgments}
We thank the referee for the careful reading of the manuscript and for the useful comments. This work was funded by the National Key R\&D Program of China (grant No.2024YFA1611504), the NSFC Grands 12403077, 12303072, 11933011, the National SKA Program of China (grant No. 2022SKA0120103), the Xinjiang Talent Development Fund (No. XJRC-2025-KJ-YJ-CXPT-180), and the Key Laboratory for Radio Astronomy. 
\end{acknowledgments}
\facility{VLBA.}

\appendix

\section{Additional Astrometric Results\label{SecA}}

\setcounter{table}{0}
\renewcommand{\thetable}{A\arabic{table}}
\renewcommand{\theHtable}{A.\arabic{table}}
\setcounter{figure}{0}
\renewcommand{\thefigure}{A\arabic{figure}}
\renewcommand{\theHfigure}{A.\arabic{figure}}

\begin{figure}
    \centering
    \includegraphics[width=0.49\linewidth]{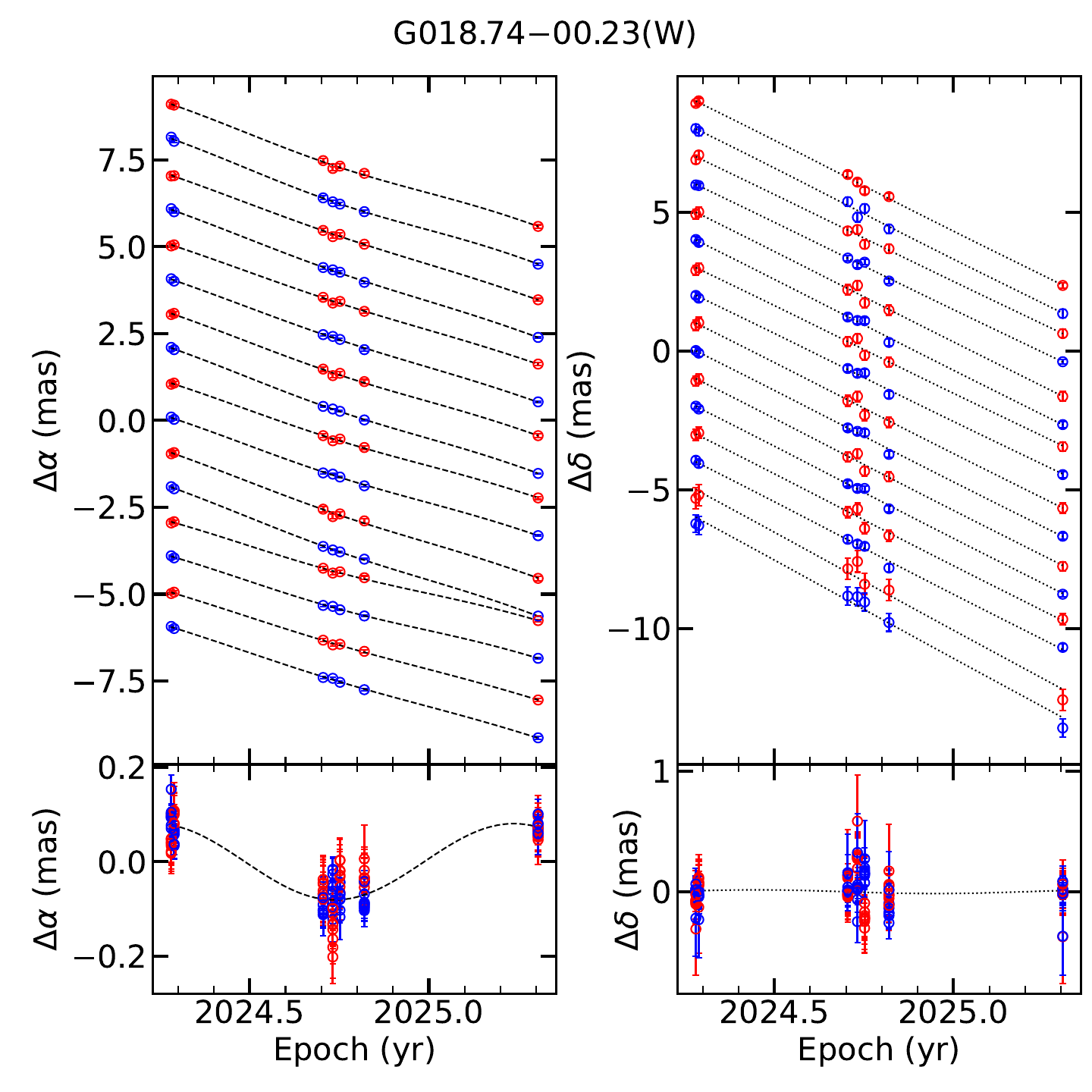}
    \includegraphics[width=0.49\linewidth]{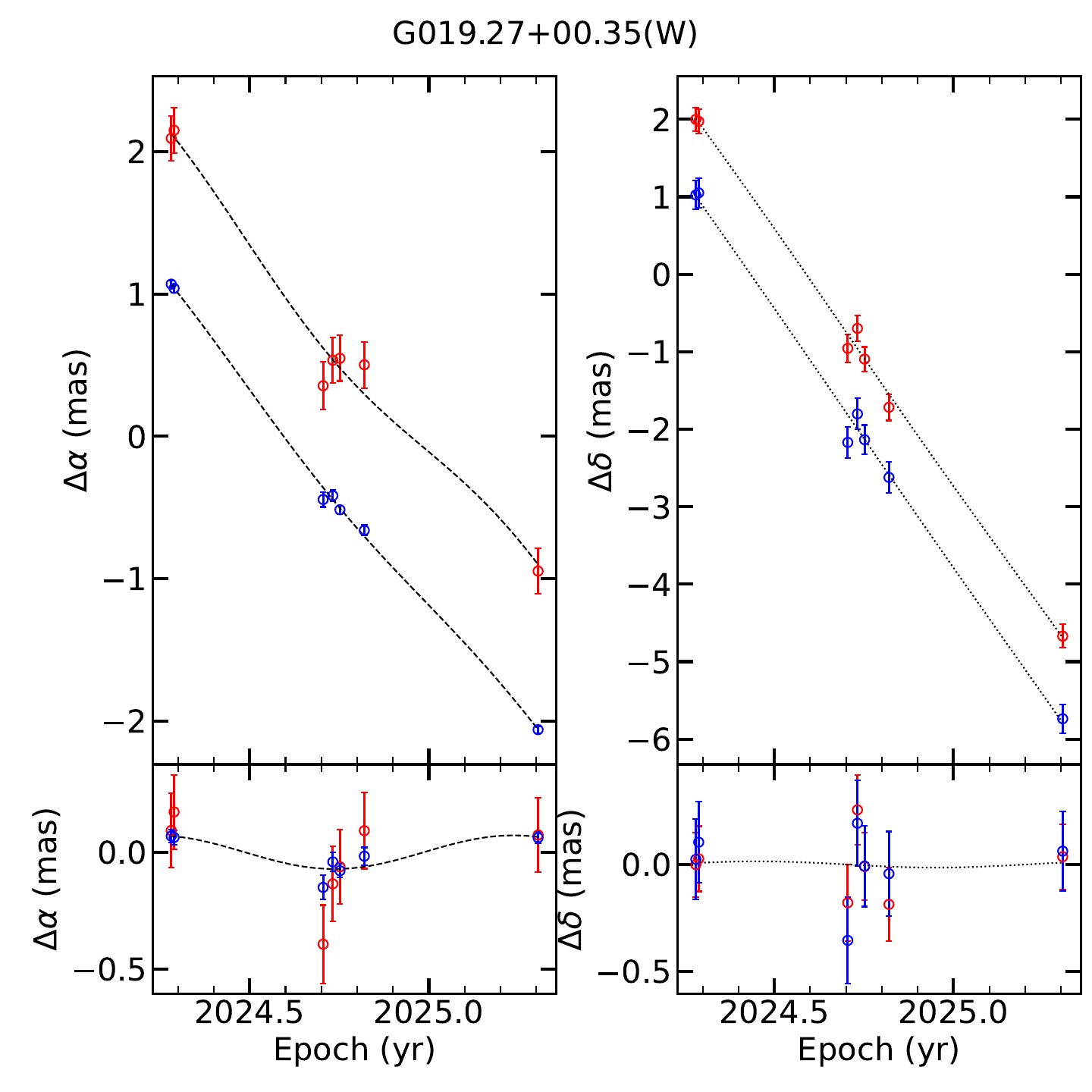}
    \includegraphics[width=0.49\linewidth]{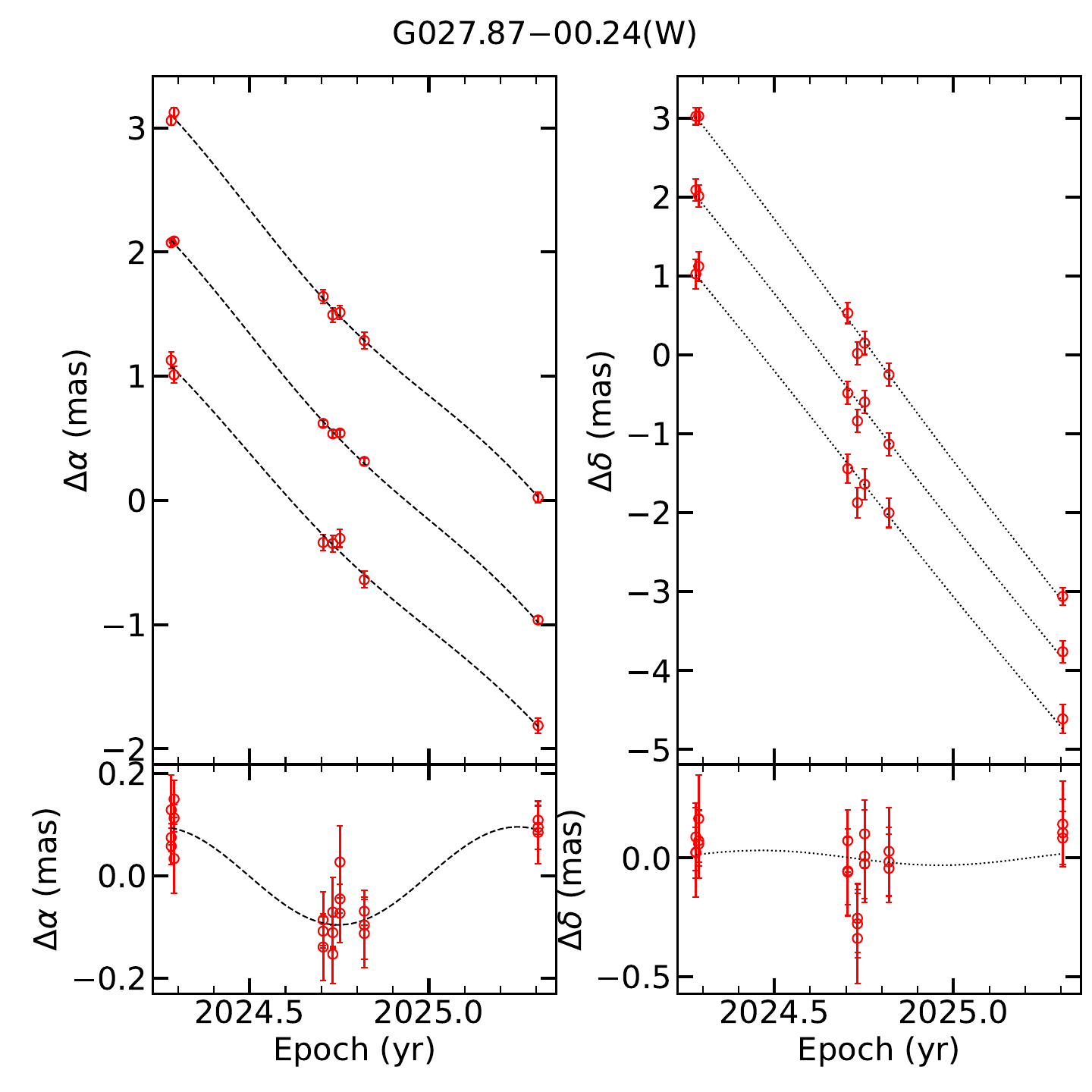}
    \includegraphics[width=0.49\linewidth]{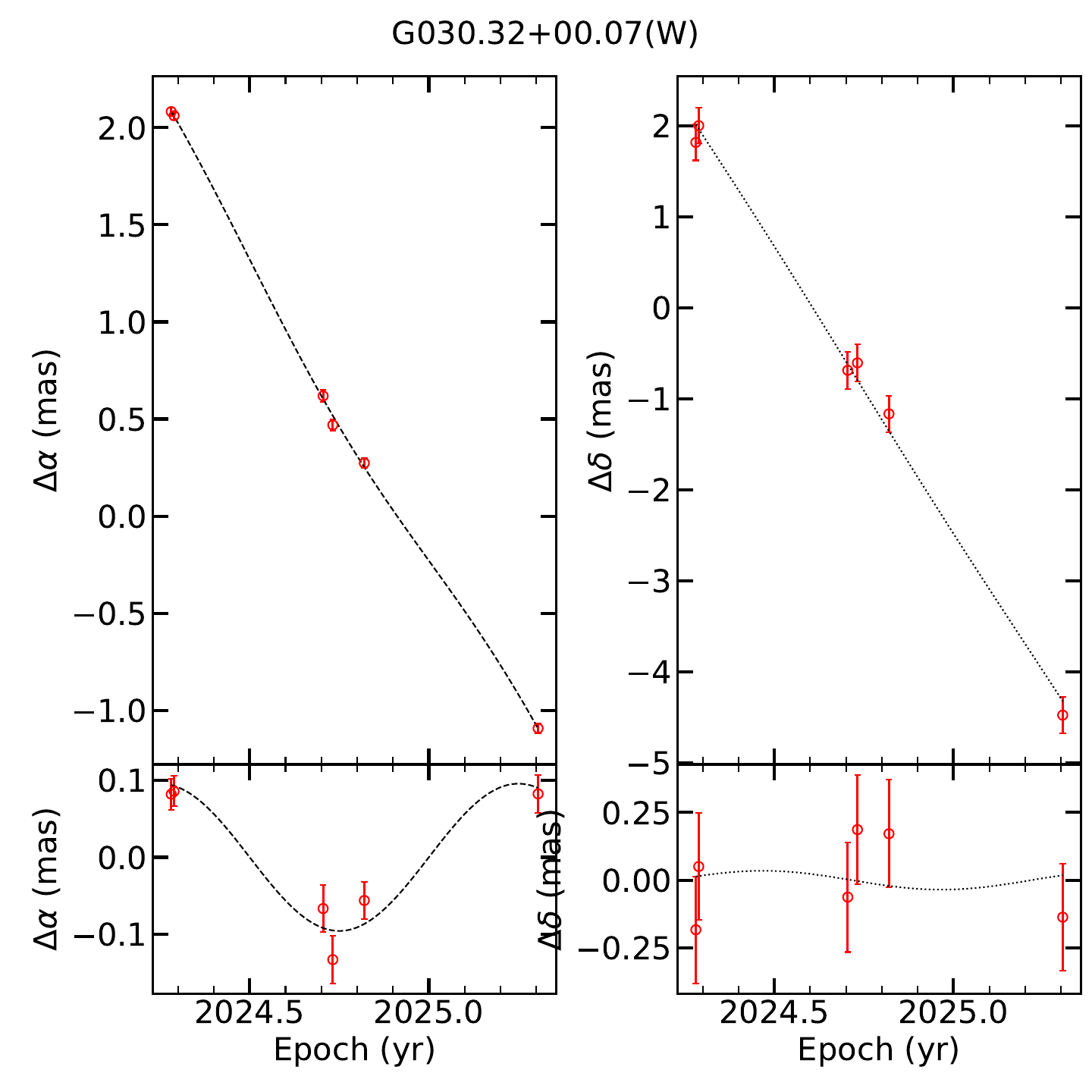}
    \caption{
    Parallax and proper-motion fits for the H$_2$O masers. The upper-left and lower-left panels show the measured right ascension offsets and the parallax signature after removing the fitted proper motion, respectively. The right panels present the corresponding results in declination. Different symbols indicate different background sources: red circles denote J1821$-$1224 for G018 and G019 and J1841$-$0348 for G027 and G030, while blue triangles denote J1818$-$1108 for G018 and G019. Different curves correspond to different maser spots included in the astrometric fitting. To display the proper motions and parallax signatures of multiple maser spots clearly in the same panel, constant positional offsets have been applied to individual maser spots; these shifts are for visualization only and do not affect the astrometric fits.}
    \label{figs:fit_water}
\end{figure}

\begin{figure}
    \centering
    \includegraphics[width=0.49\linewidth]{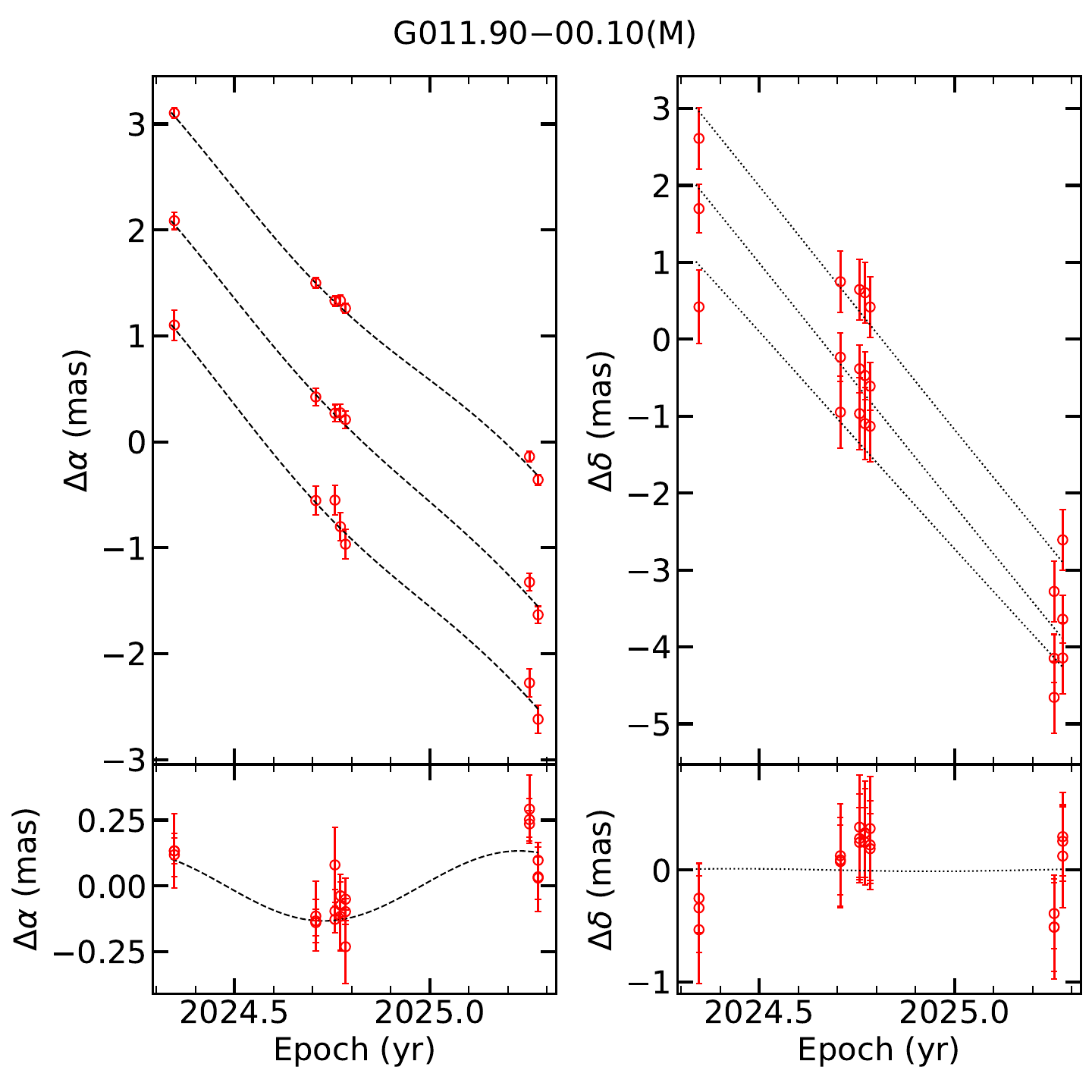}
    \includegraphics[width=0.49\linewidth]{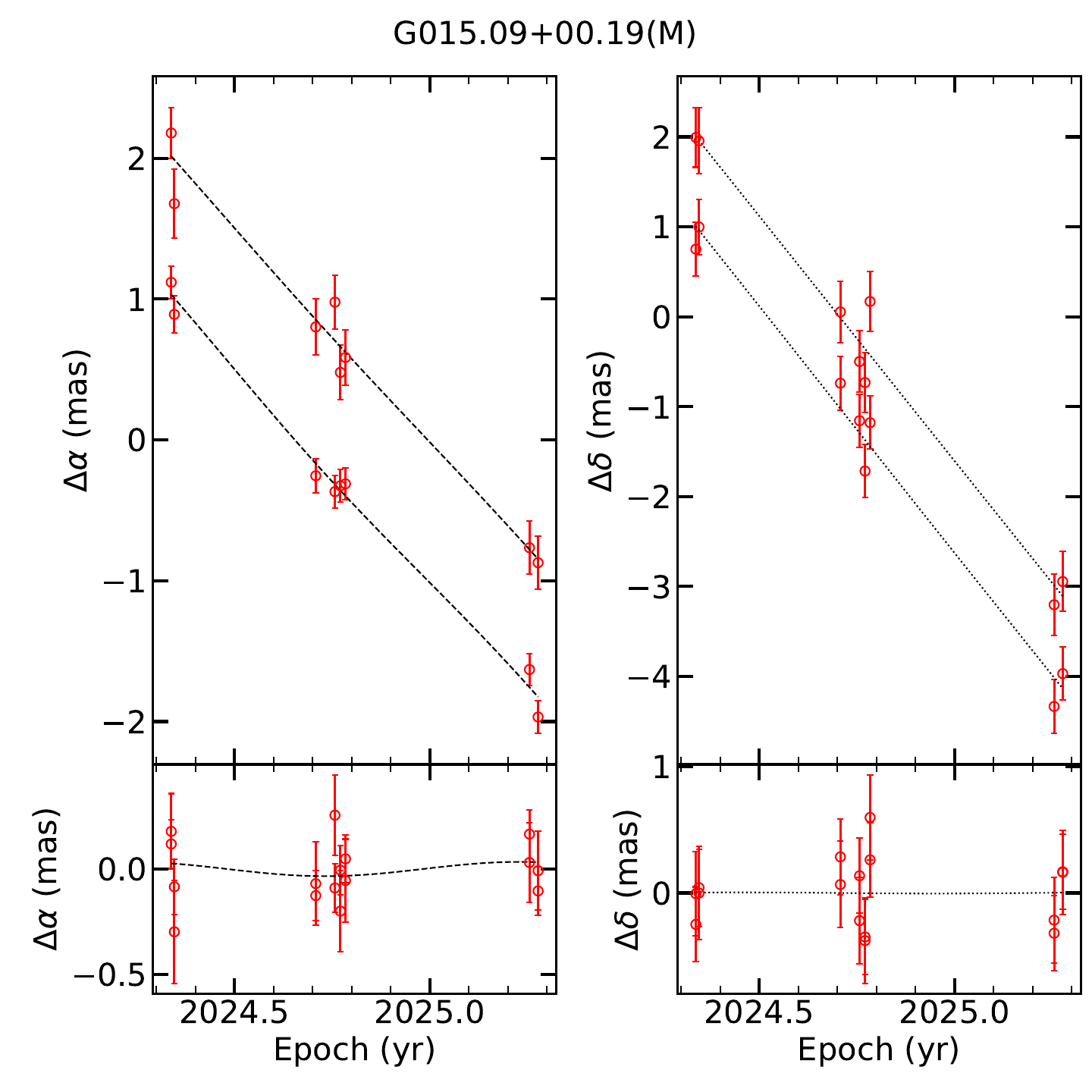}
    \includegraphics[width=0.49\linewidth]{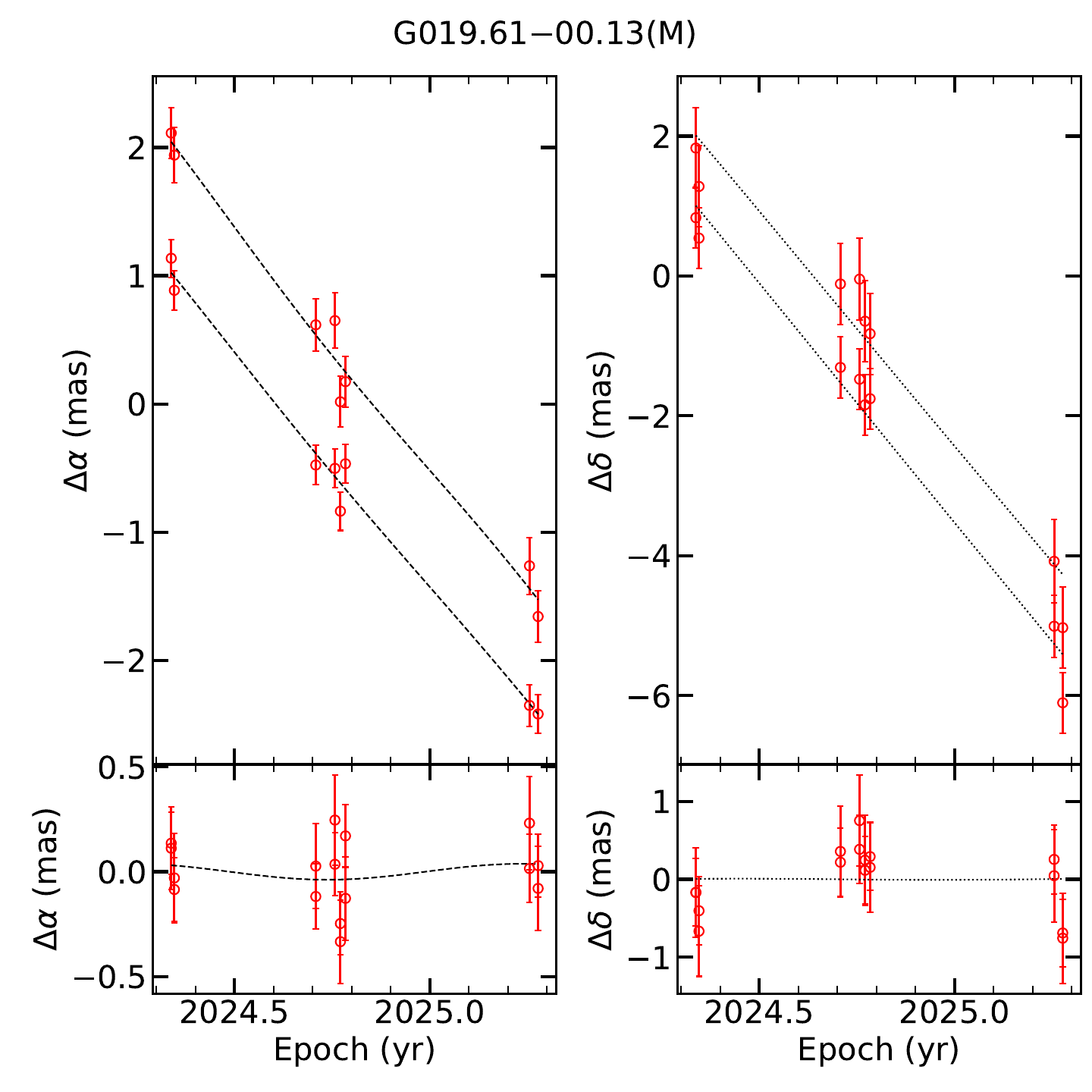}
    \includegraphics[width=0.49\linewidth]{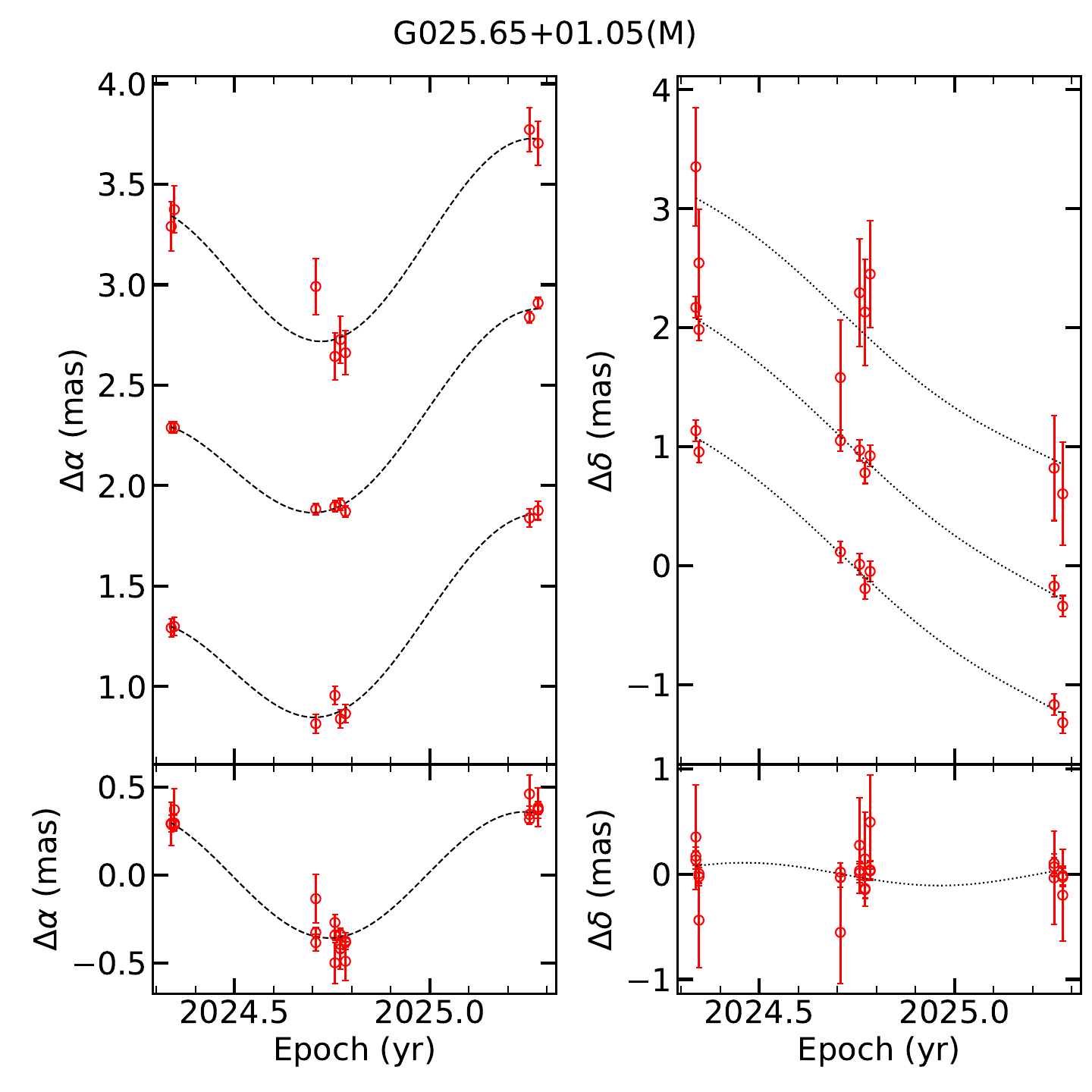}
    \caption{Parallax and proper-motion fits for the CH$_3$OH masers, presented in the same format as Figure~\ref{figs:fit_water}. Since the iMV calibration was applied to these sources, different curves represent maser spots at different velocity components used in the astrometric fitting.}
    \label{figs:fit_methanol}
\end{figure}

Here, we show the details of the epochs observed (Table~\ref{tab:epochs}),
observational parameters (Table~\ref{tab:positions_brightnesses}), and parallax fits of the
individual sources (Table~\ref{tab:parallax_pm} and Figures~\ref{figs:fit_water}$-$\ref{figs:fit_methanol}). The uncertainties of
parallaxes and proper motions given in Table~\ref{tab:parallax_pm} are the formal
fitting uncertainties.

\begin{deluxetable}{cc|c}[!ht]
\tablecaption{Details of the Epochs Observed\label{tab:epochs}}
\tablewidth{0pt}
\tablehead{
\colhead{Epoch} & 
\multicolumn{1}{c|}{Water maser} & 
\colhead{Methanol maser} \\
\colhead{} &
\multicolumn{1}{c|}{G018.74$-$00.23 \& G019.27$+$00.35} &
\colhead{G011.90$-$00.10 \& G015.09$+$00.19} \\
\colhead{} &
\multicolumn{1}{c|}{G027.87$-$00.24 \& G030.32$+$00.07} &
\colhead{G019.61$-$00.13 \& G025.65$+$01.05}
}
\startdata
E1 & 12 Apr 2024 &  3 May 2024\\
E2 & 15 Apr 2024 &  6 May 2024\\
E3 & 14 Sep 2024 & 15 Sep 2024\\
E4 & 24 Sep 2024 &  3 Oct 2024\\
E5 &  1 Oct 2024 &  8 Oct 2024\\
E6 & 26 Oct 2024 & 13 Oct 2024\\
E7 & 28 Mar 2025 &  3 Apr 2025\\
E8 & 21 Apr 2025 & 11 Apr 2025\\
\enddata
\end{deluxetable}

\begin{deluxetable*}{lcccccc}[!ht]
\tablecaption{Positions and Brightnesses\label{tab:positions_brightnesses}}
\tabletypesize{\footnotesize}
\tablewidth{0pt}
\tablehead{
\colhead{Source} &
\colhead{R.A. (J2000)} &
\colhead{Decl. (J2000)} &
\colhead{$\phi$} &
\colhead{Brightness} &
\colhead{$V_{\rm LSR}$} &
\colhead{NW beam} \\
\colhead{} &
\colhead{(h m s)} &
\colhead{($^\circ$ $'$ $''$)} &
\colhead{($^\circ$)} &
\colhead{(Jy beam$^{-1}$)} &
\colhead{(km s$^{-1}$)} &
\colhead{(mas, mas, deg)}
}
\startdata
G018.74$-$00.23(W)\tablenotemark{*} & 18 25 56.4697  & $-$12 42 49.035  & \nodata & 17.6  & 40.6  & $2.2 \times 1.3\,@\,-3$    \\
J1821$-$1224   & 18 21 23.27790 & $-$12 24 12.9348 & 1.2     & 0.019 & \nodata & $2.2 \times 1.2\,@\,2$    \\
J1818$-$1108   & 18 18 19.31374 & $-$11 08 48.3187 & 2.4     & 0.051 & \nodata & $2.6 \times 1.6\,@\,-7$    \\
G019.27$+$00.35(W)\tablenotemark{*} & 18 24 52.3769  & $-$11 58 28.033  & \nodata & 2.9  & 21.6  & $2.2 \times 1.3\,@\,1$    \\
J1821$-$1224   & 18 21 23.27790 & $-$12 24 12.9348 & 1.0     & 0.019 & \nodata & $2.3 \times 1.3\,@\,10$    \\
J1818$-$1108   & 18 18 19.31374 & $-$11 08 48.3187 & 1.8     & 0.056 & \nodata & $2.8 \times 1.7\,@\,5$    \\
G027.87$-$00.24(W)\tablenotemark{*} & 18 43 01.5583  & $-$04 36 42.588  & \nodata & 5.7  & 15.2  & $2.2 \times 1.2\,@\,-3$    \\
J1841$-$0348   & 18 41 27.31330 & $-$03 48 44.3518 & 0.9     & 0.028 & \nodata & $2.0 \times 1.4\,@\,-7$    \\
G030.32$+$00.07(W)\tablenotemark{*} & 18 46 24.9635  & $-$02 17 40.546  & \nodata & 8.7  & 47.1  & $2.8 \times 1.7\,@\,22$    \\
J1841$-$0348   & 18 41 27.31330 & $-$03 48 44.3518 & 2.0     & 0.031 & \nodata & $3.1 \times 1.6\,@\,17$    \\
G011.90$-$00.10(M)\tablenotemark{*} & 18 12 02.6822  & $-$18 40 25.002  & \nodata & 9.8  & 33.9  & $7.6 \times 4.0\,@\,12$    \\
J1808$-$1822   & 18 08 55.51545 & $-$18 22 53.3962 & 0.8     & 0.067 & \nodata & $8.3 \times 5.0\,@\,19$    \\
J1819$-$2036   & 18 19 36.89548 & $-$20 36 31.5727 & 2.7     & 0.034 & \nodata & $10.2 \times 6.0\,@\,30$    \\
G015.09$+$00.19(M) & 18 17 20.8185  & $-$15 43 46.570  & \nodata & 5.0  & 25.7  & $5.0 \times 3.6\,@\,30$    \\
J1825$-$1551   & 18 25 11.72241 & $-$15 51 34.3770 & 2.0     & 0.018 & \nodata & $5.6 \times 5.0\,@\,26$    \\
J1809$-$1520\tablenotemark{*}   & 18 09 10.20954 & $-$15 20 09.7090 & 2.1     & 0.061 & \nodata & $4.8 \times 3.6\,@\,22$    \\
G019.61$-$00.13(M)\tablenotemark{*} & 18 27 16.5225  & $-$11 53 38.156  & \nodata & 3.0  & 56.6  & $8.3 \times 5.0\,@\,34$    \\
J1818$-$1108   & 18 18 19.31374 & $-$11 08 48.3187 & 2.4     & 0.258 & \nodata & $10.1 \times 6.3\,@\,43$    \\
J1844$-$1324   & 18 44 50.28162 & $-$13 24 44.3979 & 4.7     & 0.028 & \nodata & $8.8 \times 4.1\,@\,40$    \\
G025.65$+$01.05(M)\tablenotemark{*} & 18 34 20.9023  & $-$05 59 42.196  & \nodata & 21.1  & 41.8  & $4.2 \times 1.3\,@\,-14$    \\
J1827$-$0405   & 18 27 45.04053 & $-$04 05 44.5757 & 2.5     & 0.045 & \nodata & $4.2 \times 3.0\,@\,-15$    \\
J1833$-$0323   & 18 33 23.90513 & $-$03 23 31.4450 & 2.6     & 0.043 & \nodata & $4.3 \times 2.3\,@\,-10$    \\
J1825$-$0737   & 18 25 37.60953 & $-$07 37 30.0131 & 2.7     & 0.219 & \nodata & $4.5 \times 1.8\,@\,-12$    \\
J1848$-$0822   & 18 48 13.88168 & $-$08 22 01.4170 & 4.2     & 0.032 & \nodata & $4.0 \times 2.5\,@\,7$    \\
\enddata
\tablecomments{$\phi$ is the angular separation between a maser and its calibrator. For each source, the absolute position and peak brightness of the strongest maser spot, together with the size and position angle (measured from north through east) of the naturally weighted (NW) synthesized beam, are listed for the first epoch. The maser positions and peak brightnesses were determined by Gaussian fitting with \texttt{JMFIT} (see Appendix~\ref{SecA} for details)}. $^*$ represents phase-reference. V$_{\rm LSR}$ is the velocity of the phase-reference channel or the velocity of the peak brightness.
\end{deluxetable*}

\clearpage
\startlongtable
\begin{deluxetable*}{lccccccc}
\tablecaption{Detailed Results of the Parallaxes and Proper Motions of Masers\label{tab:parallax_pm}}
\tablewidth{0pt}
\tabletypesize{\footnotesize}
\renewcommand{\arraystretch}{1.1}
\vspace{2cm}
\tablehead{
\colhead{Background} &
\colhead{$V_{\rm LSR}$} &
\colhead{Detected} &
\colhead{Parallax} &
\colhead{$\mu_x$} &
\colhead{$\mu_y$} &
\colhead{$\Delta x$} &
\colhead{$\Delta y$} \\
\colhead{Source} &
\colhead{(km s$^{-1}$)} &
\colhead{Epochs} &
\colhead{(mas)} &
\colhead{(mas y$^{-1}$)} &
\colhead{(mas y$^{-1}$)} &
\colhead{(mas)} &
\colhead{(mas)}
}
\startdata
\multicolumn{8}{c}{G018.74$-$00.23(W)} \\
\hline
J1818$-$1108 & 33.79 & 11 1111 11 & $0.106\pm0.020$ & $-3.43\pm0.05$ & $-6.48\pm0.11$ & $241.027\pm0.021$ & $146.764\pm0.033$\\
& 34.63 & 11 1111 11 & $0.057\pm0.018$ & $-3.50\pm0.05$ & $-6.23\pm0.17$ & $-601.343\pm0.025$ & $641.519\pm0.041$\\
& 38.42 & 11 1111 11 & $0.055\pm0.018$ & $-3.35\pm0.05$ & $-6.48\pm0.20$ & $-92.443\pm0.015$ & $257.983\pm0.024$ \\
& 39.58 & 11 1111 11 & $0.082\pm0.022$ & $-3.43\pm0.06$ & $-6.27\pm0.20$ & $-92.589\pm0.014$ & $276.989\pm0.023$ \\
& 40.64 & 11 1111 11 & $0.075\pm0.018$ & $-3.22\pm0.05$ & $-6.51\pm0.22$ & $-4.945\pm0.014$ & $5.700\pm0.023$ \\
& 40.95 & 11 1111 11 & $0.073\pm0.027$ & $-3.52\pm0.08$ & $-6.58\pm0.20$ & $-90.271\pm0.015$ & $293.482\pm0.024$ \\
& 42.11 & 11 1111 11 & $0.085\pm0.017$ & $-2.77\pm0.05$ & $-6.57\pm0.22$ & $36.266\pm0.015$ & $33.247\pm0.024$ \\
& 45.69 & 11 1111 11 & $0.052\pm0.018$ & $-3.02\pm0.06$ & $-7.03\pm0.44$ & $293.392\pm0.016$ & $182.309\pm0.027$ \\
\multicolumn{3}{c}{Combined fit} & $0.072\pm0.019$  \\
J1821$-$1224 & 33.79 & 11 1111 11 & $0.114\pm0.016$ & $-3.52\pm0.04$ & $ -6.49\pm0.18$ & $246.560\pm0.016$ & $140.829\pm0.026$\\
& 34.63 & 11 1111 11 & $0.070\pm0.013$ & $-3.59\pm0.03$ & $-6.25\pm0.10$ & $-595.809\pm0.022$ & $635.584\pm0.036$ \\
& 38.42 & 11 1111 11 & $0.063\pm0.013$ & $-3.45\pm0.04$ & $-6.51\pm0.13$ & $-86.909\pm0.007$ & $252.048\pm0.013$ \\
& 39.58 & 11 1111 11 & $0.089\pm0.007$ & $-3.52\pm0.02$ & $-6.29\pm0.14$ & $-87.056\pm0.007$ & $271.054\pm0.012$ \\
& 40.64 & 11 1111 11 & $0.081\pm0.011$ & $-3.31\pm0.03$ & $-6.54\pm0.13$ & $0.588\pm0.006$ & $-0.235\pm0.012$ \\
& 40.95 & 11 1111 11 & $0.082\pm0.008$ & $-3.62\pm0.02$ & $-6.59\pm0.11$ & $-84.738\pm0.007$ & $287.547\pm0.013$ \\
& 42.11 & 11 1111 11 & $0.091\pm0.009$ & $-2.86\pm0.02$ & $-6.61\pm0.15$ & $41.799\pm0.008$ & $27.312\pm0.013$ \\
& 45.69 & 11 1111 11 & $0.057\pm0.014$ & $-3.12\pm0.05$ & $-7.05\pm0.37$ & $298.925\pm0.010$ & $176.374\pm0.018$ \\
\multicolumn{3}{c}{Combined fit} & $0.083\pm0.009$  \\
Combined fit2 & & & $0.081\pm0.010$  \\
Average & & & & $-3.38\pm0.03$ & $-6.46\pm0.15$\\
\hline
\multicolumn{8}{c}{G019.27$+$00.35(W)} \\
\hline
J1818$-$1108 & 16.99 & 11 1111 11 & $0.132\pm0.066$ & $-2.95\pm0.19$ & $-6.55\pm0.18$ & $278.154\pm0.018$ & $26.559\pm0.029$ \\
J1821$-$1224 & 16.99 & 11 1111 11 & $0.069\pm0.013$ & $-3.05\pm0.03$ & $-6.64\pm0.22$ & $283.640 \pm 0.017$ & $20.604 \pm 0.027$ \\
Combined fit2 & & & $0.072\pm0.013$  \\
Average & & & & $-3.05\pm0.04$ & $-6.59\pm0.20$\\
\hline
\multicolumn{8}{c}{G027.87$-$00.24(W)} \\
\hline
J1841$-$0348 & 12.32 & 11 1111 11 & $0.106\pm0.020$ & $-3.00\pm0.05$ & $-6.00\pm0.13$ & $3.908\pm0.020$ & $13.806\pm0.035$ \\
& 15.79 & 11 1111 11 & $0.096\pm0.011$ & $-3.00\pm0.05$ & $-5.73\pm0.16$ & $-0.038\pm0.011$ & $0.404\pm0.018$ \\
& 20.74 & 11 1111 11 & $0.082\pm0.026$ & $-2.83\pm0.08$ & $-5.62\pm0.21$ & $41.396\pm0.033$ & $103.085\pm0.056$ \\
\multicolumn{3}{c}{Combined fit} & $0.096\pm0.012$  \\
Average & & & & $-2.97\pm0.04$ & $-5.84\pm0.16$\\
\hline
\multicolumn{8}{c}{G030.32$+$00.07(W)} \\
\hline
J1841$-$0348 & 12.32 & 11 1101 01 & $0.086\pm0.011$ & $-3.10\pm0.03$ & $-6.19\pm0.23$ & $-2.814\pm0.009$ & $5.313\pm0.012$ \\
\hline
\multicolumn{8}{c}{G011.90$-$00.10(M)} \\
\hline
J1808 \& J1819 & 33.91 & 11 1111 11 & $0.141\pm0.022$ & $-3.68\pm0.07$ & $-6.29\pm0.49$ & $-0.346\pm0.006$ & $-0.768\pm0.009$ \\
& 34.36 & 11 1111 11 & $0.112\pm0.036$ & $-3.90\pm0.11$ & $-6.28\pm0.38$ & $-5.780\pm0.020$ & $-3.198\pm0.030$ \\
& 35.26 & 11 1111 11 & $0.137\pm0.059$ & $-3.89\pm0.19$ & $-5.61\pm0.60$ & $-12.843\pm0.097$ & $-3.774\pm0.141$ \\
\multicolumn{3}{c}{Combined fit} & $0.133\pm0.030$  \\
Average & & & & $-3.75\pm0.09$ & $-6.15\pm0.47$\\
\hline
\multicolumn{8}{c}{G015.09$+$00.19(M)} \\
\hline
J1825 \& J1809 & 23.75 & 11 1111 11 & $0.018\pm0.077$ & $-3.05\pm0.21$ & $-5.44\pm0.36$ & $8.914\pm0.055$ & $-5.533\pm0.056$ \\
& 25.73 & 11 1111 11 & $0.040\pm0.046$ & $-3.05\pm0.13$ & $-5.47\pm0.32$ & $0.480\pm0.026$ & $0.547\pm0.025$ \\
\multicolumn{3}{c}{Combined fit} & $0.034\pm0.055$  \\
Average & & & & $-3.05\pm0.16$ & $-5.46\pm0.34$\\   
\hline
\multicolumn{8}{c}{G019.61$-$00.13(M)} \\
\hline
J1818 \& J1844 & 54.93 & 11 1111 11 & $0.054\pm0.077$ & $-3.81\pm0.22$ & $-6.68\pm0.62$ & $-2.408\pm0.058$ & $1.887\pm0.074$ \\
& 56.55 & 11 1111 11 & $0.029\pm0.058$ & $-3.67\pm0.16$ & $-6.83\pm0.47$ & $1.303\pm0.023$ & $0.399\pm0.028$ \\
\multicolumn{3}{c}{Combined fit} & $0.038\pm0.066$  \\
Average & & & & $-3.72\pm0.19$ & $-6.77\pm0.53$\\    
\hline
\multicolumn{8}{c}{G025.65$+$01.05(M)} \\
\hline
4 QSOs & 40.29 & 11 1111 11 & $0.417\pm0.045$ & $+0.34\pm0.12$ & $-2.34\pm0.49$ & $400.118\pm0.071$ & $-164.322\pm0.263$ \\
& 42.00 & 11 1111 11 & $0.354\pm0.011$ & $+0.57\pm0.03$ & $-2.48\pm0.10$ & $0.046\pm0.006$ & $-2.090\pm0.010$ \\
& 41.82 & 11 1111 11 & $0.362\pm0.019$ & $+0.54\pm0.05$ & $-2.44\pm0.10$ & $-0.196\pm0.005$ & $-2.253\pm0.007$ \\
\multicolumn{3}{c}{Combined fit} & $0.359\pm0.016$  \\
Average & & & & $+0.55\pm0.04$ & $-2.46\pm0.12$\\   
\enddata
\tablecomments{
$\Delta x$ and $\Delta y$ are the positional offsets measured at the first epoch, using different background QSOs and/or different maser velocity channels, relative to the corresponding reference positions listed in Table~\ref{tab:positions_brightnesses}. The parallax, $\mu_x$, and $\mu_y$ were obtained using the astrometric fitting procedure described in Appendix~\ref{SecA}. ``Combined fit'' denotes a simultaneous fit to multiple maser spots at different velocities, with a common parallax fitted to all spots. ``Combined fit 2'' further combines the parallax measurements obtained from all maser spots and different background QSOs. Because systematic astrometric errors can be correlated among maser spots observed at the same epoch, the formal uncertainty of a combined parallax was conservatively increased by a factor of $\sqrt{N_{\rm spot}}$, where $N_{\rm spot}$ is the number of maser spots included in the fit. ``Average'' gives the weighted averages of the individual proper-motion measurements, with the quoted uncertainties likewise accounting for the $\sqrt{N_{\rm spot}}$ factor.}
\end{deluxetable*}

\clearpage
\section{Far-side Maser Sources in the Perseus and Sagittarius Arms}

Here, we summarize the $V_{\rm LSR}$ and distances of the far-side maser sources associated with the Perseus and Sagittarius Arms.

\setcounter{table}{0}
\renewcommand{\thetable}{B\arabic{table}}
\renewcommand{\theHtable}{B.\arabic{table}}

\begin{deluxetable}{lcclcclcc}[!ht]
\tablecaption{Far-side Maser Sources in the Perseus and Sagittarius Arms
\label{tab:masers_all}}
\tablewidth{0pt}
\tablehead{
\\
\colhead{Maser Source} &
\colhead{$V_{\rm LSR}$} &
\colhead{Distance} &
\colhead{Maser Source} &
\colhead{$V_{\rm LSR}$} &
\colhead{Distance} &
\colhead{Maser Source} &
\colhead{$V_{\rm LSR}$} &
\colhead{Distance} \\
\colhead{} &
\colhead{(km s$^{-1}$)} &
\colhead{(kpc)} &
\colhead{} &
\colhead{(km s$^{-1}$)} &
\colhead{(kpc)} &
\colhead{} &
\colhead{(km s$^{-1}$)} &
\colhead{(kpc)}
}
\startdata
\multicolumn{9}{c}{Sagittarus Arm}\\
\hline
G011.90$-$00.10$^*$ & 34$\pm$5  &  12.9$\pm$1.1 & G018.74$-$00.23$^*$ & 38$\pm$5  &  12.8$\pm$0.7 & G019.60$-$00.23 & 41$\pm$3  & 12.6$\pm$0.7 \\
G019.61$-$00.13$^*$ & 55$\pm$5  &  12.0$\pm$1.0 & G020.08$-$00.13 & 41$\pm$3  &  12.7$\pm$0.6 & G030.32$+$00.07$^*$ & 45$\pm$5  &  11.7$\pm$0.8 \\
G032.74$-$00.07 & 37$\pm$10 &  10.6$\pm$0.8 & G033.64$-$00.22 & 61$\pm$3  &   9.9$\pm$0.5 & G035.57$-$00.03 & 53$\pm$3  &  10.2$\pm$0.6 \\
G035.79$-$00.17 & 61$\pm$5  &   9.4$\pm$0.6 & G037.47$-$00.10 & 58$\pm$3  &   9.6$\pm$0.9 & G038.03$-$00.30 & 60$\pm$3  &   9.3$\pm$0.6 \\
G041.15$-$00.20 & 60$\pm$3  &   7.6$\pm$0.5 & G041.22$-$00.19 & 59$\pm$5  &   8.7$\pm$1.1 & G043.03$-$00.45 & 56$\pm$5  &   8.2$\pm$0.6 \\
G043.89$-$00.78 & 54$\pm$3  &  7.5$\pm$0.3 & G045.07$+$00.13 & 59$\pm$5  &   7.7$\pm$0.4 & G045.45$+$00.06 & 55$\pm$7  &   8.1$\pm$0.9 \\
G045.49$+$00.12 & 58$\pm$3  &   6.9$\pm$0.9 & G045.80$-$00.35 & 64$\pm$5  &   7.0$\pm$1.0 \\
\hline
\multicolumn{9}{c}{Perseus Arm}\\
\hline
G001.00$-$00.23 &  2$\pm$5  & 13.9$\pm$ 3.4 & G015.09$+$00.19$^*$ & 26$\pm$5  & 15.1$\pm$ 1.4 & G019.27$+$00.35$^*$ & 16$\pm$5  & 13.2$\pm$ 0.8\\
G021.87$+$00.01 & 19$\pm$10 & 13.5$\pm$ 1.2 & G027.87$-$00.24$^*$ & 20$\pm$5  & 12.7$\pm$ 1.0 & G031.24$-$00.11 & 24$\pm$10 & 13.1$\pm$ 0.8 \\
G032.79$+$00.19 & 16$\pm$10 & 12.5$\pm$ 1.2 & G037.49$+$00.52 & 11$\pm$10 & 12.5$\pm$ 0.7 & G037.82$+$00.41 & 16$\pm$10 & 12.1$\pm$ 0.6 \\
G040.42$+$00.70 & 10$\pm$5  & 12.1$\pm$ 0.4 & G040.62$-$00.13 & 31$\pm$5  & 10.8$\pm$ 0.9 & G042.03$+$00.19 & 12$\pm$5  & 11.9$\pm$ 0.5 \\
G043.16$+$00.01 & 11$\pm$5  & 11.6$\pm$ 0.5 & G048.60$+$00.02 & 18$\pm$5  & 10.3$\pm$ 0.4 & G049.26$+$00.31 &  0$\pm$5  & 10.5$\pm$ 1.0 \\
G049.41$+$00.32 &$-12\pm$5  & 11.6$\pm$ 0.9 & G050.28$-$00.39 & 17$\pm$5  &  9.4$\pm$ 0.8 & G060.57$-$00.18 &  3$\pm$5  &  7.7$\pm$ 0.4 \\
G070.18$+$01.74 &$-23\pm$5  &  7.5$\pm$ 0.5 & G070.29$+$01.60 &$-27\pm$10 &  9.1$\pm$ 0.5 & G070.32$+$01.58 &$-23\pm$5  &  8.1$\pm$ 1.2 \\
\enddata
\tablecomments{$^{*}$ indicates sources newly reported in this work. The remaining Sagittarius Arm sources are taken from \citet{bian2024}, while the Perseus Arm sources are taken from \citet{hyland2026}; references to the original maser astrometric measurements can be found in those works.}
\end{deluxetable}

\end{CJK*}

\begin{thebibliography}{aasjournal}

\bibitem[Bartkiewicz et al.(2011)]{bartkiewicz2011} Bartkiewicz, A., Szymczak, M., Pihlstr{\"o}m, Y.~M., et al.\ 2011, \aap, 525, A120. doi:10.1051/0004-6361/201015235

\bibitem[Bayandina et al.(2019)]{bayandina2019} Bayandina, O.~S., Burns, R.~A., Kurtz, S.~E., et al.\ 2019, \apj, 884, 2, 140. doi:10.3847/1538-4357/ab3fa4

\bibitem[Bian et al.(2024)]{bian2024} Bian, S.~B., Wu, Y.~W., Xu, Y., et al.\ 2024, \aj, 167, 6, 267. doi:10.3847/1538-3881/ad4030


\bibitem[Breen et al.(2015)]{breen2015} Breen, S.~L., Fuller, G.~A., Caswell, J.~L., et al.\ 2015, \mnras, 450, 4, 4109. doi:10.1093/mnras/stv847

\bibitem[Brunthaler et al.(2011)]{brunthaler2011} Brunthaler, A., Reid, M.~J., Menten, K.~M., et al.\ 2011, Astronomische Nachrichten, 332, 5, 461. doi:10.1002/asna.201111560

\bibitem[Burns et al.(2020)]{burns2020} Burns, R.~A., Orosz, G., Bayandina, O., et al.\ 2020, \mnras, 491, 3, 4069. doi:10.1093/mnras/stz3172

\bibitem[Cohen et al.(1980)]{cohen1980} Cohen, R.~S., Cong, H., Dame, T.~M., et al.\ 1980, \apjl, 239, L53. doi:10.1086/183290


\bibitem[Dame et al.(2001)]{dame2001} Dame, T.~M., Hartmann, D., \& Thaddeus, P.\ 2001, \apj, 547, 2, 792. doi:10.1086/318388

\bibitem[Green \& McClure-Griffiths(2011)]{green2011} Green, J.~A. \& McClure-Griffiths, N.~M.\ 2011, \mnras, 417, 4, 2500. doi:10.1111/j.1365-2966.2011.19418.x

\bibitem[Hyland et al.(2022)]{hyland2022} Hyland, L.~J., Reid, M.~J., Ellingsen, S.~P., et al.\ 2022, \apj, 932, 1, 52. doi:10.3847/1538-4357/ac6d5b

\bibitem[Hyland et al.(2023)]{hyland2023} Hyland, L.~J., Reid, M.~J., Orosz, G., et al.\ 2023, \apj, 953, 1, 21. doi:10.3847/1538-4357/acdbc5

\bibitem[Hyland et al.(2026)]{hyland2026} Hyland, L.~J., Reid, M.~J., Ellingsen, S.~P., et al.\ 2026, \apj, 1004, 2, 209. doi:10.3847/1538-4357/ae64f5

\bibitem[Molinari et al.(1996)]{molinari1996} Molinari, S., Brand, J., Cesaroni, R., et al.\ 1996, \aap, 308, 573. 

\bibitem[Mookerjea \& Ghosh(1999)]{mookerjea1999} Mookerjea, B. \& Ghosh, S.~K.\ 1999, Journal of Astrophysics and Astronomy, 20, 1-2, 1. doi:10.1007/BF02715036

\bibitem[Moscadelli et al.(2002)]{moscadelli2002} Moscadelli, L., Menten, K.~M., Walmsley, C.~M., et al.\ 2002, \apj, 564, 2, 813. doi:10.1086/324304

\bibitem[Reid et al.(2009)]{reid2009} Reid, M.~J., Menten, K.~M., Zheng, X.~W., et al.\ 2009, \apj, 700, 1, 137. doi:10.1088/0004-637X/700/1/137

\bibitem[Reid et al.(2016)]{reid2016} Reid, M.~J., Dame, T.~M., Menten, K.~M., et al.\ 2016, \apj, 823, 2, 77. doi:10.3847/0004-637X/823/2/77

\bibitem[Reid et al.(2017)]{reid2017} Reid, M.~J., Brunthaler, A., Menten, K.~M., et al.\ 2017, \aj, 154, 2, 63. doi:10.3847/1538-3881/aa7850

\bibitem[Reid et al.(2019)]{reid2019} Reid, M.~J., Menten, K.~M., Brunthaler, A., et al.\ 2019, \apj, 885, 2, 131. doi:10.3847/1538-4357/ab4a11

\bibitem[Reid (2022)]{reid2022} Reid, M.~J.\ 2022, \aj, 164, 4, 133. doi:10.3847/1538-3881/ac80bb

\bibitem[Rioja et al.(2017)]{rioja2017} Rioja, M.~J., Dodson, R., Orosz, G., et al.\ 2017, \aj, 153, 3, 105. doi:10.3847/1538-3881/153/3/105

\bibitem[Sakai et al.(2023)]{sakai2023} Sakai, N., Zhang, B., Xu, S., et al.\ 2023, \pasj, 75, 1, 208. doi:10.1093/pasj/psac102

\bibitem[Sunada et al.(2007)]{sunada2007} Sunada, K., Nakazato, T., Ikeda, N., et al.\ 2007, \pasj, 59, 1185. doi:10.1093/pasj/59.6.1185

\bibitem[Titmarsh et al.(2014)]{titmarsh2014} Titmarsh, A.~M., Ellingsen, S.~P., Breen, S.~L., et al.\ 2014, \mnras, 443, 4, 2923. doi:10.1093/mnras/stu1346

\bibitem[VERA Collaboration et al.(2020)]{vera2020} VERA Collaboration, Hirota, T., Nagayama, T., et al.\ 2020, \pasj, 72, 4, 50. doi:10.1093/pasj/psaa018

\bibitem[Volvach et al.(2019)]{volvach2019} Volvach, L.~N., Volvach, A.~E., Larionov, M.~G., et al.\ 2019, \mnras, 482, 1, L90. doi:10.1093/mnrasl/sly193

\bibitem[Weaver(1970)]{weaver1970} Weaver, H.\ 1970, The Spiral Structure of our Galaxy, 38, 126. 

\bibitem[Xu et al.(2006)]{xu2006} Xu, Y., Reid, M.~J., Zheng, X.~W., et al.\ 2006, Science, 311, 5757, 54. doi:10.1126/science.1120914

\bibitem[Xu et al.(2021)]{xu2021apjs} Xu, Y., Bian, S.~B., Reid, M.~J., et al.\ 2021, \apjs, 253, 1, 1. doi:10.3847/1538-4365/abd8cf

\bibitem[Xu et al.(2023)]{xu2023} Xu, Y., Hao, C.~J., Liu, D.~J., et al.\ 2023, \apj, 947, 2, 54. doi:10.3847/1538-4357/acc45c

\end{thebibliography}
\end{document}